\documentclass[a4paper,12pt]{article}
\usepackage[top=0.7in, bottom=1.0in, left=1.0in, right=1.0in]{geometry}
\usepackage{amsmath,soul,multicol}
\usepackage{graphicx,stfloats,pdflscape,color,hyperref}
\usepackage[onehalfspacing]{setspace}
\usepackage[T1]{fontenc}
\usepackage[utf8]{inputenc}
\usepackage{mathptmx}
\usepackage{lineno}
\usepackage{amsmath}

\begin{document}

\setstretch{1.0}

\noindent {\huge \bf Extended main sequence turn-off originated 
from a broad range of stellar rotational velocities } \\ 

\noindent {\bf Beomdu Lim$^{1,2,*}$ Gregor Rauw$^1$, Ya\"el Naz\'e$^1$, Hwankyung Sung$^2$, Narae Hwang$^3$, and Byeong-Gon Park$^{3,4}$} \\

\noindent $^1$Space sciences, Technologies and Astrophysics Research (STAR) Institute, 
                  Universit\'e de Li\`ege, Quartier Agora, All\'ee du 6 Ao\^ut 19c, B\^at. B5C, 4000, Li\`ege, Belgium\\
\noindent $^2$Department of Physics and Astronomy, Sejong University, 209 Neungdong-ro, Gwangjin-gu, Seoul 05006, Republic of Korea\\
\noindent $^3$School of Space Research, Kyung Hee University, 1732 Deogyeong-daero, Giheung-gu, Yongin-si, Gyeonggi-do 17104, Republic of Korea\\
\noindent $^4$Korea Astronomy and Space Science Institute, 776 Daedeokdae-ro, Yuseong-gu, Daejeon 34055, Republic of Korea\\
\noindent $^5$Astronomy and Space Science Major, University of Science and Technology, 217 Gajeong-ro, Yuseong-gu, Daejeon 34113, Republic of Korea\\

\noindent $^*$Corresponding author (blim@uliege.be)\\

{\bf \noindent  Star clusters have long been considered to comprise a simple stellar 
population, but this paradigm is being challenged, since apart from multiple populations 
in Galactic globular clusters$^{1,2}$, a number of intermediate-age star clusters exhibit a 
significant colour spread at the main sequence turn-off (MSTO)$^{3,4,5,6,7,8,9,10,11}$. A 
sequential evolution of multiple generations of stars formed over 100-200 million years 
is a natural explanation of this colour spread$^{12}$. Another approach to explain this 
feature is to introduce the effect of stellar rotation$^{13}$. However, its effectiveness has 
not yet been proven due to the lack of direct measurements of rotational velocities. 
Here we report the distribution of projected rotational velocities ($V\sin i$) of stars in the 
Galactic open cluster M11, measured through a Fourier transform analysis. Cluster 
members display a broad $V\sin i$ distribution, and fast rotators including Be stars have 
redder colours than slow rotators. Monte Carlo simulations infer that cluster members 
have highly aligned spin axes and a broad distribution of equatorial velocities biased towards 
high velocities. Our findings demonstrate how stellar rotation affects the colours of cluster 
members, suggesting that the colour spread observed in populous clusters can be understood 
in the context of stellar evolution even without introducing multiple stellar populations.  }\\

The total mass of star clusters ranges from about $10^2 M_{\odot}$ to $\sim 10^6 M_{\odot}$ 
(Ref. 14). The minimum initial cluster mass required to retain material 
ejected from the first generation of stars is about $10^4 M_{\odot}$ (Ref. 15), and 
therefore most open clusters ($< 10^4 M_{\odot}$) are expected to have had 
insufficient initial mass to host multiple generations of stars with different abundances. 
This is confirmed by the fact that many open clusters show a negligible variation in terms of 
chemical composition$^{16,17}$ among cluster members. But surprisingly, a colour 
spread of stars at the MSTO, the so-called ``extended MSTO'' feature, similar to the one 
discovered in populous clusters of the Magellanic Clouds (MCs)$^{3,4,5,6,7,8,9}$, was 
found in the Hyades and Praesepe$^{10,11}$ open clusters. The suspicion therefore 
remains of the presence of multiple populations in open clusters. However, stellar 
rotation has also been advanced to explain the existence of an extended MSTO,  
and it is important to determine which explanation is correct. To this aim, Galactic open 
clusters would be ideal testbeds because their proximity permits the precise derivation of 
stellar properties, necessary to clarify the origin of the extended MSTO.

M11 is an intermediate-age cluster (250 Myr) at 2 kpc distance$^{18}$, having 
a total mass up to 11000 $M_{\odot}$ (Refs. 19 and 20). Red clump stars in the cluster 
show a slight enhancement of $\alpha$-elements with respect to the solar abundance, 
but they have a homogeneous chemical composition$^{20,21,22}$. Fig. 1 shows the 
colour-magnitude diagram of stars in the cluster region (see Methods). In contrast to the 
narrow distribution in V of the subgiant branch stars, the cluster members 
near the MSTO ($V = 12.5$ mag) are spread over a wide colour range of 0.3 mag. 
Given the narrower main sequence band in the fainter regime, photometric 
errors and a high binary fraction cannot explain the observed spread near the MSTO. If the 
age spread scenario$^{12}$ is applied to the MSTO, the observed colour spread 
corresponds to an age spread of about 150 Myr. This age spread is comparable to those 
found in populous clusters in the MCs$^{3,4,5,7,9,12}$. 

We analyzed high-resolution optical spectra of 164 cluster members near the 
MSTO obtained from {\it Gaia}-European Southern Observatory (ESO) Public 
Spectroscopic Survey internal Data Release 4$^{23,24}$ and from observations 
with {\it Hectochelle} on the 6.5m Multiple Mirror Telescope (see Methods for 
details). The red spectra of 164 stars covering the H$\alpha$ line were used to identify 
Be stars, while blue spectra of 108 stars containing several weak metallic lines 
as well as the main spectral line Mg {\scriptsize \textsc{II}} $\lambda 4481$ 
were used in the determination of $V\sin i$.

The presence of rapidly rotating stars in populous clusters of the MCs were 
inferred through the identification of Be stars$^{25,26}$ either from 
H$\alpha$ photometry or from limited spectroscopy due to their large 
distance. But, it remains to be seen whether their colours are different 
from those of slow rotators or not. In this work, we identified a total of 22 
Be stars in M11 by visually inspecting the $H\alpha$ line profile (Supplementary 
Fig. 1). Their H$\alpha$ emission line was found to have a double-peaked shell 
profile, which is indicative of absorption by a circumstellar disc along the line 
of sight. We directly measured $V\sin i$ of 108 stars, including the identified 
Be stars, with available blue spectra by applying a Fourier transform technique$^{27}$ 
to the Mg {\scriptsize \textsc{II}} $\lambda 4481$ line for $V\sin i \geq 25$ 
km s$^{-1}$, or to the other metallic lines for slower rotators (see Methods for 
further explanation). 

Fig. 2 compares the colours and $V\sin i$ of cluster members. The distribution 
of very fast rotators ($V\sin i > 200$ km s$^{-1}$) is extended towards redder colours, 
while the distribution of slow rotators is shifted towards bluer colours (Fig. 2b). 
Fig 2c further shows the correlation between the two measurements with a correlation 
coefficient of about 44 per cent. According to this correlation, a rotating star with 
$V\sin i =$ 300 km s$^{-1}$ has a $U-V$ colour 0.15 mag redder than a non-rotating 
star. This finding suggests that stellar rotation plays a significant role in the colour 
spread at the MSTO. We also confirmed that Be stars in M11 are indeed rapidly rotating 
stars (see also Supplementary Fig. 2) and tend to have colours redder than those of slowly 
rotating stars. 

Rotational mixing plays a similar role to that of convective overshooting in the mixing of 
internal material. Fresh hydrogen is supplied to the core through the mixing 
process, and therefore the main sequence lifetime of a rotating star is prolonged 
by 15 -- 62 per cent for a given mass and metallicity$^{28}$. If cluster members 
have a broad distribution of rotational velocities, the main sequence band is 
expected to broaden in colour-magnitude diagrams due to their different main sequence 
lifetimes. Fast rotation also deforms the shape of a star, lowering the equatorial 
temperature and luminosity as the equator expands. This gravity darkening effect 
causes spreads in both magnitude and colour, depending on the inclination angles of 
spin axes with respect to an observer. In order to test the impact of rotation on the 
colour-magnitude diagram of the cluster members, the underlying distributions 
of equatorial velocities ($V_{\mathrm{eq}}$) and inclination angles ($i$) need to 
be constrained. We introduced four combinations of $V_{\mathrm{eq}}$ and $i$ 
distributions to infer their underlying distributions. Case 1 considers a uniform 
distribution of $V_{\mathrm{eq}}$ and a uniform orientation of $i$ in a 
three-dimensional (3D) space, Case 2 treats a uniform distribution of $V_{\mathrm{eq}}$ 
and a Gaussian distribution of $i$, Case 3 adopts a linear distribution of $V_{\mathrm{eq}}$ 
and a uniform orientation of $i$ in a 3D space, and Case 4 uses a linear 
distribution of $V_{\mathrm{eq}}$ and a Gaussian distribution of $i$ (see Methods 
for details). There is no free parameter for Case 1; a peak inclination angle 
($i_{\mathrm{peak}}$) and a dispersion ($\sigma_{i}$) are the 
free parameters to be determined for Case 2; for Case 3, the slope $(\alpha)$ of 
the linear distribution is the only free parameter; the free parameters 
$i_{\mathrm{peak}}$, $\sigma_i$, and $\alpha$ are required to obtain the 
distributions of $V_{\mathrm{eq}}$ and $i$ for Case 4. We derived those free 
parameters by comparing the distribution of the observed $V\sin i$ to that 
of the simulated $V\sin i$ based on the chosen $V_{\mathrm{eq}}$ and $i$ 
distributions (see Methods for further explanation, Supplementary Table 1). 

Fig. 3. displays the cumulative distributions of both the observed and simulated 
$V\sin i$. Case 1 simulations yielded distributions very different from the observed 
one since the mean probability that the observed distribution is drawn from the 
simulated case is only about $6.2 \pm 4.0$ (s.d.) per cent according to the 
Kolmogorov-Smirnov (K-S) test. Adopting the best-fit parameters ($i_{\mathrm{peak}} 
= 54^{\circ}$ and $\sigma_i = 1^{\circ}$), Case 2 simulations reproduce the observed 
$V\sin i$ distribution with a mean confidence level of $42.1 \pm 13.2$ per cent. For 
Case 3, we found a best-fit parameter of $\alpha = 0.0006$ km$^{-1}$ s, however 
the similarity of the two distributions is only about $10.7 \pm 4.6$ per cent. Finally, 
Case 4 simulations yield the best match to the observed distribution with a mean 
confidence level of $74.1 \pm 12.8$ per cent; the derived parameters 
$i_{\mathrm{peak}}$, $\sigma_i$, and $\alpha$ are 50$^{\circ}$, 2$^{\circ}$, and 
0.0013 km$^{-1}$ s, respectively. This result means that the best-fit distribution of 
$i$ is a Gaussian distribution centered on 50$^{\circ}$ with a sigma of only 2$^{\circ}$: that 
implies a strong alignment of spin axes.  If the collapsing molecular cloud is rotating, 
and if it has a rotational kinetic energy comparable to the turbulent kinetic energy, 
the spin axes of stars formed in the cloud can be aligned. The gravitational interaction 
among cluster members does not significantly change the spin axes, and therefore 
the spin alignment can remain for several billion years$^{29}$. The strong alignment 
of spin axes in M11 can be understood in this context. In addition, the derived slope 
of the linear distribution of $V_{\mathrm{eq}}$ indicates that cluster members may 
have a distribution of $V_{\mathrm{eq}}$ biased towards high rotational velocities, rather 
than a uniform distribution.
 
Adopting the $V_{\mathrm{eq}}$ and $i$ distributions inferred from Case 4 
simulations, a synthetic cluster was generated by taking into account rotational 
velocity, gravity darkening, the photometric errors of the used data$^{18}$, the 
observed differential reddening$^{18}$, and a minimum binary fraction of 5 
per cent (see Methods for further explanation). The colour-magnitude diagram 
of M11 and the synthetic cluster are compared in Fig. 4, demonstrating the 
similarity between these clusters (especially the colour-rotation spread). In our simulation (Case 4), 
gravity darkening impacts effective temperature and luminosity by $1.5\pm0.4$ and 
$7.6\pm3.4$ per cent, respectively, on average, which correspond to impacts of 
only $-0.021\pm 0.006$ mag on the $U-V$ colours and $-0.080\pm 0.035$ on 
the bolometric magnitude for an A0-type star (Supplementary Fig. 3). On the other 
hand, the differential reddening in M11 accounts for 0.03 mag (1$\sigma$) 
in $B-V^{18}$, which corresponds to 0.05 mag (1$\sigma$) in $U-V$ according to the usual 
reddening law$^{30}$. Since the mean $U-V$ photometric error is about 0.02 mag 
and the total spread is about 0.07 mag ($1\sigma$), the colour spread due to the 
rotation is about 0.05 mag $\Big(\sigma_{\mathrm{rot}} = \sqrt{\sigma^2_{\mathrm{obs}} - \sigma^2_{\mathrm{red}} - \sigma^2_{\mathrm{err}}}\Big)$: differential reddening 
and rotation thus have comparable effect on the $U-V$ colour spread.

Our findings thus indicate that the observed extended MSTO can result from a 
broad distribution of stellar rotational velocities and does not require the presence 
of multiple populations. The absence of subgiant branch stars connecting to the 
red hook of the MSTO also strongly supports a single stellar population scenario 
for M11 as seen in the populous cluster NGC 1651$^{10}$. The direct comparison 
between the colours and $V\sin i$ of Be stars proved that their reddened colours 
are due to fast rotation. This conclusion should certainly be extended to the populous 
clusters in the MCs. Our discovery finally challenges ideas about cluster evolution as 
such clusters were often considered as young counterparts of globular clusters, which 
comprise multiple populations$^{1,2}$, incompatible with our result.\\

\noindent{\bf References} \\
\noindent 1. Hesser, J. E., Hartwick, F. D. A. \& McClure, R. D. Cyanogen strengths and ultraviolet excesses of evolved stars in 17 globular clusters from DDO photometry.  {\it Astrophys. J. Suppl. Ser.} {\bf 33,} 471-493 (1977)\\
2. Milone, A. P. {\it et al.} Multiple Stellar Populations in 47 Tucanae. {\it Astrophys. J.} {\bf 744,} 58-79 (2012) \\
3. Bertelli, G. {\it et al.} Testing intermediate-age stellar evolution models with VLT photometry of Large Magellanic Cloud clusters. III. Padova results. {\it Astron. J.} {\bf 125,} 770-784 (2003).\\
4. Mackey, A. D. \& Nielsen, P. B. A double main-sequence turn-off in the rich star cluster NGC 1846 in the
Large Magellanic Cloud. {\it Mon. Not. R. Astron. Soc.} {\bf 379,} 151-158 (2007)\\
5. Goudfrooij, P. {\it et al.} Population parameters of intermediate-age star clusters in the Large Magellanic Cloud. II. New insights from extended main-sequence turnoff in seven star clusters. {\it Astrophys. J.} {\bf 737,} 3-20 (2011)\\
6. Girardi, L. {\it et al.} An extended main-sequence turn-off in the Small Magellanic Cloud star cluster NGC 411. {\it Mon. Not. R. Astron. Soc.} {\bf 431,} 3501-3509 (2013)\\
7. Li, C., de Grijs, R. \& Deng, L. Not-so-simple stellar populations in the intermediate-age Large Magellanic Cloud star clusters NGC 1831 and NGC 1868. {\it Astrophys. J.} {\bf 784,} 157-169 (2014)\\
8. Li, C., de Grijs, R. \& Deng, L. The exclusion of a significant range of ages in a massive star cluster. {\it Nature} {\bf 516,} 367-369 (2014)\\
9. Milone, A. P. {\it et al.} Multiple stellar populations in Magellanic Cloud clusters - III. The first evidence of an extended main sequence turn-off in a young cluster: NGC 1856. {\it Mon. Not. R. Astron. Soc.} {\bf 450,} 3750-3764 (2015)\\
10. Eggen, O. J. The age range of Hyades stars. {\it Astron. J.} {\bf 116,} 284-292 (1998)\\
11. Brandt, T. D. \& Huang, C. X. The age and age spread of the Praesepe and Hyades clusters: A consistent, $\sim$800 Myr picture from rotating stellar models. {\it Astrophys. J.} {\bf 807,} 24-29 (2015)\\
12. Goudfrooij, P. {\it et al.} Population parameters of intermediate-age star clusters in the Large Magellanic Cloud. I. NGC 1846 and its wide main-sequence turnoff. {\it Astron. J.} {\bf 137,} 4988-5002 (2009)\\
13. Bastian, N. \& de Mink, S. E. The effect of stellar rotation on colour–magnitude diagrams: on the apparent presence of multiple populations in intermediate age stellar clusters. {\it Mon. Not. R. Astron. Soc.} {\bf 398,} L11–L15 (2009).\\
14. Turner, J. L. Extreme Star Formation. {\it Astrophysics and Space Science Proceedings} 10, 215
(2009). 1009.1416.\\
15. Conroy, C. \& Spergel, D. N. On the formation of multiple stellar populations in globular clusters. {\it Astrophys. J.} {\bf 726,} 36-48 (2011)\\
16. Pancino, E. {\it et al.} Chemical abundance analysis of the open clusters Cr 110, NGC 2099 (M 37), NGC 2420, NGC 7789, and M 67 (NGC 2682). {\it Astron. Astrophys.} {\bf 511,} 56-74 (2010)\\
17. Reddy, A. B. S., Giridhar, S. \& Lambert, D. L. Comprehensive abundance analysis of red giants in the open clusters NGC 2527, 2682, 2482, 2539, 2335, 2251 and 2266. {\it Mon. Not. R. Astron. Soc.} {\bf 431,} 3338-3348 (2013)\\
18. Sung, H. {\it et al.} $UBVI$ CCD photometry of M11 - II. New photometry and surface density profiles, {\it Mon. Not. R. Astron. Soc.} {\bf 310}, 982-1001 (1999)\\
19. Santos Jr. J. F. C., Bonatto, C. \& Bica, E. Structure and stellar content analysis of the open cluster M 11 with 2MASS photometry. {\it Astron. Astrophys.} {\bf 442,} 201-209 (2005)\\
20. Cantat-Gaudin, T. {\it et al.} The Gaia-ESO Survey: Stellar content and elemental abundances in the massive cluster NGC 6705. {\it Astron. Astrophys.} {\bf 569,} 17-34 (2014)\\
21. Magrini, L. {\it et al.} The Gaia-ESO Survey: Abundance ratios in the inner-disk open clusters Trumpler 20, NGC 4815, NGC 6705. {\it Astron. Astrophys.} {\bf 563,} 44-57 (2014)\\
22. Tautvai$\check{\mathrm{s}}$ien$\dot{\mathrm{e}}$, G. {\it et al.} The Gaia-ESO Survey: CNO abundances in the open clusters Trumpler 20, NGC 4815, and NGC 6705. {\it Astron. Astrophys.} {\bf 573,} 55-67 (2015) \\
23. Gilmore, G. {\it et al.} The Gaia-ESO Public Spectroscopic Survey, {\it The Messenger} {\bf 147}, 25-31 (2012)\\
24. Randich, S., Gilmore, G., \& Gaia-ESO Consortium. The Gaia-ESO Large Public Spectroscopic Survey, {\it The Messenger} {\bf 514}, 47-49 (2013) \\
25. Milone, A. P. {\it et al.} Multiple stellar populations in Magellanic Cloud clusters – VI. A survey of multiple sequences and Be stars in young clusters. {\it Mon. Not. R. Astron. Soc.} {\bf 477,} 2640–2663 (2018) \\
26. Dupree, A. K. {\it et al.} NGC 1866: First Spectroscopic Detection of Fast-rotating Stars in a Young LMC Cluster.  {\it Astrophys. J. Lett.} {\bf 846,} 1-7 (2017)\\
27. Gray, D. F. {\it The observation and analysis of stellar photospheres}, {\bf 3rd ed.} (Cambridge University Press), 458-504 (2005)\\
28. Georgy, C. {\it et al.} Populations of rotating stars I. Models from 1.7 to 15 $M_{\odot}$ at $Z = 0.014$, 0.006, and 0.002 with $\Omega / \Omega_{\mathrm{cri}}$ between 0 and 1. {\it Astron. Astrophys.} {\bf 553,} 24-40 (2013)\\
29. Corsaro E. {\it et al.} Spin alignment of stars in old open clusters, {\it Nature Astron.} {\bf 1}, 0064-    (2017)\\
30. Guetter, H. H. \& Vrba, F. J. Reddening and polarimetric studies toward IC 1805. {\it Astron. J.} {\bf 98,} 611-746 (1989)\\

\clearpage

\begin{figure}
\centering
\includegraphics[angle=0,width=80.0mm]{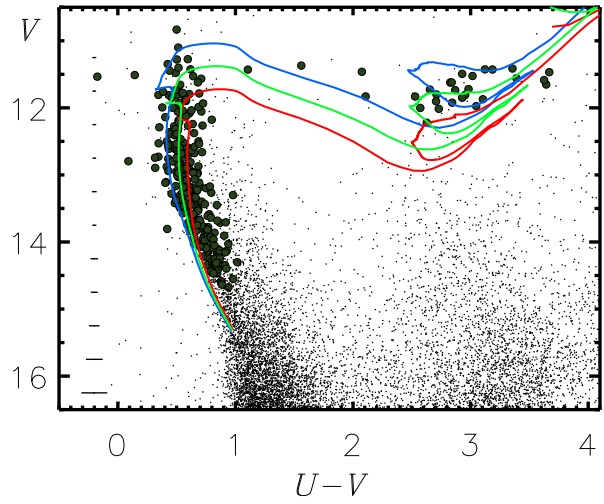}
\caption{{\bf Colour-magnitude diagram of the Galactic populous cluster M11.} 
Bold dots represent the cluster members selected from proper motion data. Error 
bars indicate mean photometric errors within given magnitude bins. Solid 
lines are isochrones of the Geneva stellar evolution models for the relative angular 
velocity to the critical value $\omega=0.7$ and $\log t = 8.4$, 8.5, and 8.6, 
respectively [yr]$^{28}$. }
\vspace{0mm} %% add extra space ONLY when figures/tables are "colliding"!
\end{figure}

\begin{figure}
\centering
\includegraphics[angle=0,width=130.0mm]{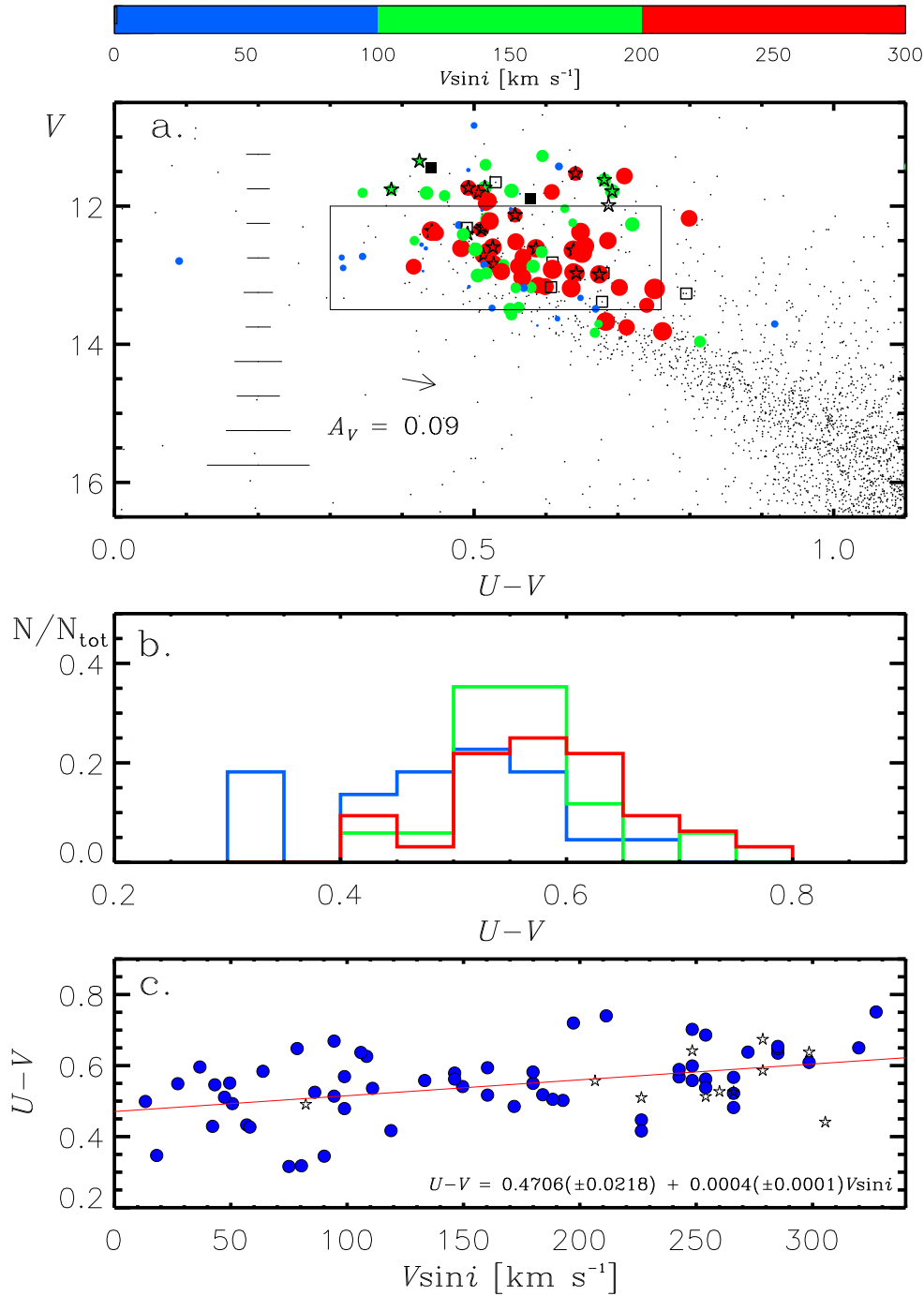}
\caption{{\bf Correlation between $V\sin i$ and $U-V$ colours.} {\bf a.} The 
colour-magnitude diagram of intermediate-mass stars near the MSTO. The size of dots 
are proportional to $V\sin i$. Star symbols, filled squares, open squares, and small dots 
represent Be stars, eclipsing binaries, double-lined spectroscopic binary candidates, 
and the other stars without $V\sin i$ measurement, respectively. Error bars display 
mean photometric errors within given magnitude bins. {\bf b.} The $U-V$ distributions 
of stars within the outlined box in panel {\bf a} in three different $V\sin i$ ranges 
that we used (identified by 3 colours, see the scale bar). Histograms were normalized 
by the total number of stars within given $V\sin i$ ranges. Note that binary stars were 
not used in this analysis. {\bf c.} The colour variation with respect to $V\sin i$ for the 
same sample as used in panel {\bf b}. Be stars are plotted as star symbols. The solid 
line represents the result of a linear least square regression, and its solution is expressed 
in the bottom of the panel. }
\vspace{0mm} %% add extra space ONLY when figures/tables are "colliding"!
\end{figure}

\newpage
\begin{figure}
\centering
\includegraphics[angle=0,width=80.0mm]{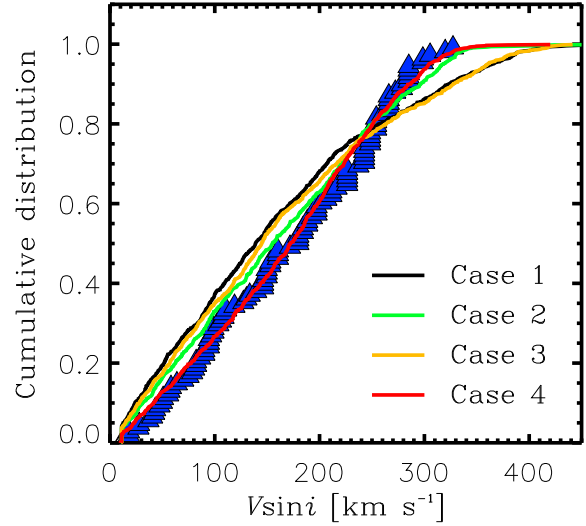}
\caption{{\bf Cumulative distribution of $V\sin i$ from observations and simulations.} Blue 
triangles represent the observed $V\sin i$ distribution, while black, green, orange, and red solid 
lines show the samples from the simulated distributions adopting the best parameters for 
Cases 1, 2, 3, and 4, respectively.}
\vspace{0mm} %% add extra space ONLY when figures/tables are "colliding"!
\end{figure}

\begin{figure}
\centering
\includegraphics[angle=0,width=80.0mm]{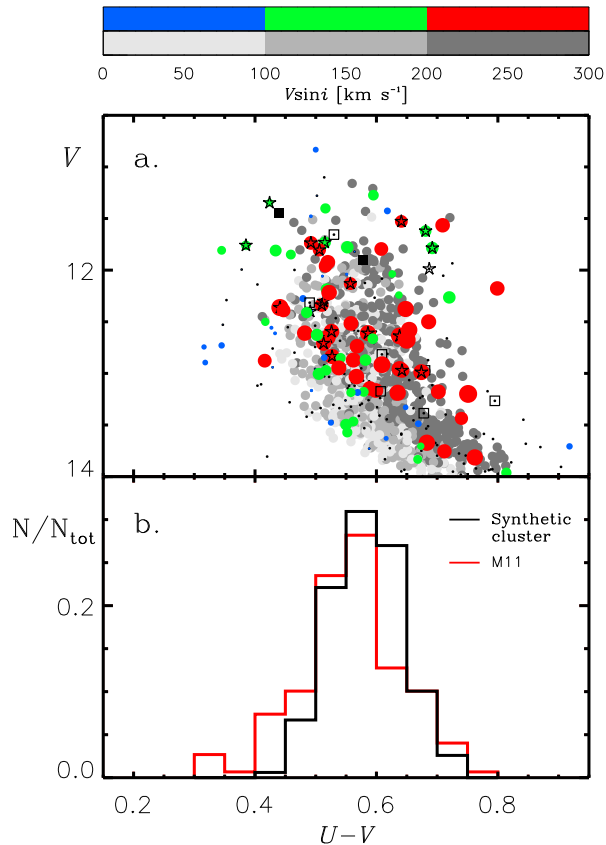}
\caption{{\bf Comparison of M11 and a synthetic cluster (single population).} 
{\bf a.} The $(V, U-V)$ colour-magnitude diagram observed for M11 (coloured dot 
as in Fig. 1.) and of our synthetic cluster (gray dot, with darken shades indicating 
faster rotating stars). {\bf b.} Colour distribution of the observed (red) and artificial 
(black) stars within the same colours and magnitude ranges as outlined in Fig. 2a. Note 
their very good match.}
\vspace{0mm} %% add extra space ONLY when figures/tables are "colliding"!
\end{figure}

\clearpage

\newpage

\noindent{\large \bf Methods}\\

\noindent{\bf Data acquisition.} The published$^{18}$ $UBV$ photometric 
data of M11 cover a $40^{\prime}\times40^{\prime}$ region centered on the 
cluster. Mean photometric errors down to the visual 
magnitude of 14 mag are about 1\% in $V$ magnitude, $V-I$, and $B-V$, and better than 
2\% in $U-B$. The proper motion of stars allows us to distinguish the bona fide 
members from field interlopers because the cluster members have almost 
the same kinematic properties.  For this purpose, we used two catalogues of 
proper motion$^{31,32}$ for stars in the M11 field in member selection. The members were selected 
using the following criteria: 1) membership probabilities higher than 70 per cent from 
both catalogues$^{31,32}$, 2) $B-V$ bluer than 0.62, and 3) visual magnitude 
brighter than 14 mag. The membership validity was tested using the proper motion 
data from {\it Gaia} Data Release 2$^{33,34}$. The proper motions of the stars 
having counterparts from the {\it Gaia} catalogue are well constrained within 
a small kinematic boundary (Supplementary Fig. 4). A total of 277 members near the
MSTO were finally included in our target list. 

Queue scheduled observations of 37 of these cluster members were carried out on 
2017 May 5, 24, and June 9 with the multi-object high-resolution spectrograph 
Hectochelle$^{35}$ attached to the 6.5-m telescope of the MMT observatory. The 
spectral resolving power of Hectochelle is about $R \sim 34000$. The observations 
of the stars were made with the order-separating filter OB 25 transmitting light in 
the wavelength range of 6475\AA \ -- 6630\AA \ in a $2\times1$ binning mode. 
Dome flat and ThAr lamp spectra were also taken, just before and after the target 
exposure. Data reduction was made according to the standard procedure for extraction 
of one dimensional spectra$^{36,37}$. Sky-subtracted spectra were combined into a 
single spectrum for the same star, and then normalized by using continuum levels 
found from a cubic spline interpolation.

The high-resolution spectra of 155 cluster members were obtained from the 
{\it Gaia}-ESO Public Spectroscopic Survey internal Data Release 4\footnote{The 
{\it Gaia}-ESO public data can be accessed through the link https://www.gaia-eso.eu/data-products/public-data-releases}(Ref. 23 and 24). Note that 28 stars were observed at both the MMT 
and Paranal observatories. The spectroscopic observations were made with 
either GIRAFFE ($R \sim 17000$ -- 26000)$^{38}$ or Ultraviolet and Visual 
Echelle Spectrograph (UVES, $R \sim 47000$)$^{39}$ attached to the 8.2m 
Very Large Telescope (Unit 2). The red spectra of 155 stars were taken with 
more than one instrumental setup among GIRAFFE/HR14A, HR15N, and UVES/U580, 
while the blue spectra of 133 stars were acquired by using either GIRAFFE/HR5A 
or UVES/U520 setups. The procedures of pre-processing and calibrations were 
made through their data reduction pipelines (see the ESO Phase 3 Data Release Description\footnote{http://www.eso.org/rm/api/v1/public/releaseDescriptions/92} 
for detail). The flattened, wavelength-corrected, and sky-subtracted spectra of given 
stars are delivered as the final data. We normalized the reduced spectra 
using the same method as above. \\

\noindent{\bf Binary stars.} We found seven double-lined spectroscopic binary 
candidates from visual inspection of the spectra. Five of them had been 
identified as spectroscopic binaries using a cross-correlation technique in 
a previous study of the same data$^{40}$. Since we are studying only 60 per cent 
of the full sample of cluster members, about a dozen spectroscopic binaries are 
expected for the full sample. In addition, two eclipsing binary members were 
found in a previous study$^{41}$. Therefore, the minimum binary fraction among 
cluster members is about 5 per cent.\\

\noindent{\bf Fourier analysis.} We used the Fourier technique to determine 
$V\sin i$ of stars$^{27}$. The profile of a spectral line can be expressed 
by a convolution of a flux profile with various broadening functions related to 
stellar rotation, thermal motion, microturbulence, macroturbulence, and 
instrumental characteristics$^{27}$. The profile of a spectral line is dominated 
by the rotation profile if the contribution of the rotational broadening is large 
enough compared to that of the other broadening sources. The rotation-dominated 
profile have zeros at low frequency in the Fourier space, while the other broadening 
components do not have zeros. The frequency 
of the first zero depends on $V\sin i$, having a relation as below:

\begin{equation}
{\lambda \over c}\sigma_1 V  \sin i  = 0.660
\end{equation}
\noindent where $\lambda$, $c$, and $\sigma_1$ represent the rest-frame 
wavelength of a given spectral line, the speed of light, and the frequency of 
the first zero. 

Our sample stars are intermediate-mass (late-B to A-type) stars, known to 
have a broad range of rotational velocities$^{42,43}$. Careful selection of 
isolated lines is important to obtain reliable $V\sin i$ because spectral lines in 
their spectra could be easily blended with adjacent lines due to the large 
rotational broadening. The Mg {\scriptsize \textsc{II}} doublet at $\lambda4481$ \AA \ is a 
moderately strong line which is not much sensitive to temperature. This line is properly 
separated from adjacent lines, and is therefore the most suitable line to 
determine $V\sin i$ for rapidly rotating A- to late-B-type stars ($V\sin i > 
100$ km s$^{-1}$). In addition, several weak metallic lines can be used in 
the $V\sin i$ range of 10 km s$^{-1}$ to 70 km s$^{-1}$.

The spectra taken with the GIRAFFE/HR5A grating contain Mg {\scriptsize \textsc{II}} 
$\lambda4481$ and a handful of metallic lines around H$\gamma$. UVES spectra 
contain more metallic lines because of its wide wavelength coverage (Supplementary 
Table 2). Those additional metallic lines were not blended with adjacent lines 
and were not seriously affected by the H$\gamma$ absorption line$^{44}$. We computed the 
Fourier transform of these lines and measured the frequency of the first zero$^{27}$. $V\sin i$ 
derived from Mg {\scriptsize \textsc{II}} $\lambda 4481$ were compared with 
those found from the other metallic lines (Supplementary Fig. 5). Since the intrinsic 
separation between the doublet components of Mg {\scriptsize \textsc{II}} 
$\lambda 4481$ dominantly influences the line profile of slow rotators ($V\sin i 
< 25$ km s$^{-1}$), the Fourier method overestimate 
$V\sin i$$^{44}$ in that case. We adopted the mean values of $V\sin i$ derived 
from the other metallic lines as the $V\sin i$ for slow rotators.  In other cases, $V\sin i$ 
derived from Mg {\scriptsize \textsc{II}} $\lambda 4481$ was consistent with those 
from the other metallic lines in the $V\sin i$ range of 25 km s$^{-1}$ to 40 km 
s$^{-1}$. Fe and Ti lines are blended with adjacent lines in the higher velocity 
regime ($V\sin i > 70$ km s$^{-1}$), and thereby yielding higher $V \sin i$ 
than those from Mg {\scriptsize \textsc{II}} $\lambda 4481$.

$V\sin i$ could be derived for 108 out of 133 stars with available blue spectra. 
The spectra of the remaining stars had either insufficient signal-to-noise ratio (SNR) to identify Mg 
{\scriptsize \textsc{II}} $\lambda4481$ or/and metallic emission lines, or they 
were eclipsing binaries. In addition, a limb darkening coefficient was also 
estimated to be 0.53 in this procedure from comparison with a pure rotation 
profile in the Fourier space$^{43}$ (Supplementary Fig. 6). This value is in a 
good agreement with a previous measurement at 4000-4500 \AA$^{45}$.  \\

\noindent{\bf Error estimation.} We estimated the errors of the measured $V\sin i$ 
through Monte-Carlo simulations. To this aim, we used synthetic spectra without 
noise for effective temperatures of 10000 K and 13000 K generated using the 
spectrum analysis code \textsc{SPECTRUM v2.76}$^{27}$ and Kurucz ODFNEW 
model atmospheres$^{46}$. Rotational broadening functions for various $V\sin i$ 
from 25 km s$^{-1}$ to 350 km s$^{-1}$ were applied to the synthetic spectra 
using the \textsc{AVSINI} programme of \textsc{SPECTRUM}, where a limb darkening 
coefficient of 0.53 was adopted. These spectra were convolved with an instrumental 
broadening function ($\Delta\lambda=0.24$\AA) using \textsc{SMOOTH2}$^{27}$. 
We carried out the Fourier transform of Mg {\scriptsize \textsc{II}} $\lambda 4481$ in the 
synthetic spectra and compared the derived $V\sin i$ with input values ($V\sin i _{\mathrm{input}}$) (Supplementary Fig. 7). 

Linear least-square fitting to the difference between $V\sin i$ and $V\sin i_{\mathrm{input}}$ 
($\Delta V\sin i$) was made in the $V\sin i_{\mathrm{input}}$ range of 0 km s$^{-1}$ to 
230 km s$^{-1}$, where the intercept was set to 0 km s$^{-1}$. We obtained $\Delta V\sin i 
= 0.03 V\sin i_{\mathrm{input}}$. This relation indicated that a systematic error of about 
3 per cent is involved in our measurements. $\Delta V\sin i$ for $V\sin i_{\mathrm{input}} 
> 230$ km s$^{-1}$ is approximated by the equation $\Delta V\sin i = 0.26 
(V\sin i_{\mathrm{input}}-230) + 7.47$. The systematic errors continuously increased 
up to 11 per cent at 350 km s$^{-1}$ as Mg {\scriptsize \textsc{II}} $\lambda 4481$ 
begins to be blended with distant He {\scriptsize \textsc{I}} $\lambda4471$ and metallic 
lines.

Noise corresponding to a given SNR was added to the synthetic 
spectra adopting $V \sin i$ of 25 km s$^{-1}$ to 350 km s$^{-1}$, respectively, using the \textsc{RANDOMU} function of \textsc{IDL}, where the SNRs were set from 10 to 100 
with an interval of 10 for a given $V\sin i$, respectively. Each simulation was repeated 100 
times for the same setup, and the $V\sin i$ were estimated in each case as made above. 
The standard deviations of the derived $V\sin i$ distributions were adopted as the random 
errors for a given SNR and $V\sin i$ (Supplementary Fig. 7). The random errors exceeded
50 km s$^{-1}$ only for SNRs lower than 20 and $V\sin i$ in the range 250 km $^{-1}$ 
to 350 km s$^{-1}$. Using those simulations, we estimated the random errors of $V\sin i$ 
for the observed SNRs of our spectra and the derived $V\sin i$ after correcting for the 
systematic errors. The errors were smaller than 30 km s$^{-1}$ in most cases (Supplementary 
Fig. 7). The typical random errors for very slow rotators ($V\sin i < 25$ km s$^{-1}$) 
was estimated to be about 1.4 km s$^{-1}$ from the standard deviation of their $V\sin i$ 
measurements.

A- and B-type stars, in general, do not have convective currents in their photospheres, 
and therefore it is difficult to imagine macroscopic motions such as the solar granules. 
Nevertheless, measurable radial-tangential macroturbulence (about 5 km s$^{-1}$) 
was found in very slowly rotating stars ($V\sin i <$ 20 km $s^{-1}$)$^{47}$. Such 
macroturbulent broadening, together with instrumental broadening, can systematically 
influence their $V\sin i$ measurements. Our sample contains four very slow rotators 
($V\sin i <$ 25 km s$^{-1}$) observed with GIRAFFE/HR5A. Since Mg {\scriptsize 
\textsc{II}} $\lambda4481$ cannot be used for them because of its doublet nature, we 
analyzed the other metallic lines (Fe {\scriptsize \textsc{I}} $\lambda 4404$, Fe 
{\scriptsize \textsc{II}} $\lambda 4489$ $\lambda 4491$ Ti {\scriptsize \textsc{II}} 
$\lambda 4468$, and $\lambda 4488$) to investigate systematic errors and the lower 
limit of measurable $V\sin i$ for the instrumental setup. 

Synthetic spectra without noise were generated using the same method as above, and 
then broadened by adopting a radial-tangential macroturbulent velocity of 5 km 
s$^{-1}$ with \textsc{MACTURB}$^{27}$. Rotational broadening functions corresponding 
to $V\sin i$ from 5 km s$^{-1}$ to 25 km s$^{-1}$ were applied to five synthetic 
spectra, respectively. Each spectrum was finally broadened with \textsc{SMOOTH2} 
according to the spectral resolution of GIRAFFE/HR5A. We then derived $V\sin i$ from 
those Fe and Ti lines. Comparison of the input $V\sin i$ with the measured ones showed 
the mean difference in the $V\sin i$ range of 10 km s$^{-1}$ to 25 km s$^{-1}$ to be 
about 1.1 km s$^{-1}$. This is the systematic error in the $V\sin i$ range. 
For the input $V\sin i$ of 5 km s$^{-1}$, the measured $V\sin i$ was overestimated 
on average by 5.4 km s$^{-1}$. Therefore, the lower limit of our measurements 
that are not affected by macroturbulent and instrumental broadening is 
about 10 km s$^{-1}$, and all the very slowly rotating stars in our sample have 
$V\sin i$ higher than the limit. \\

\noindent{\bf Underlying distribution of $V_{\mathrm{eq}}$ and $i$.} With stellar 
mass, $V_{\mathrm{eq}}$ and $i$ are the key parameters in generating 
the colour-magnitude diagram of a synthetic cluster. However, the underlying 
distributions of these parameters for the members of M11 are unknown. Here, 
we introduce a method to infer these underlying distributions using a Monte-Carlo 
technique. 

A total of 1000 artificial stars were used in each simulation. Given the masses 
of the stars at the MSTO of M11, stellar masses were generated in the range of 
2.4$M_{\odot}$ to 3.6 $M_{\odot}$, where the mass function of M11 
($\Gamma = -2.0$)$^{18}$ was used as the probability function. Stars 
can rotate up to a critical velocity ($V_{\mathrm{cri}}$) when the centrifugal 
force equals the gravitational force, and $V_{\mathrm{cri}}$ varies as a function 
of stellar mass. The maximum $V_{\mathrm{eq}}$ of given stars were taken by 
interpolating their masses to the mass-$V_{\mathrm{cri}}$ relation of the Geneva 
isochrone adopting $\omega = 0.95$ and $\log t = 8.4$ [yr]$^{28}$. Individual 
artificial stars have $V_{\mathrm{eq}}$ and $i$ in the ranges of 0 km s$^{-1}$ to 
$V_{\mathrm{cri}}$ and of 0$^{\circ}$ to 90$^{\circ}$, respectively.

We considered four cases for the probability distributions of $V_{\mathrm{eq}}$ and $i$. 
For $V_{\mathrm{eq}}$, either a uniform distribution or a linear distribution was 
adopted, while either a uniform orientation in 3D space or a Gaussian distribution 
was assumed for $i$. The uniform distribution of $V_{\mathrm{eq}}$ 
means that $V_{\mathrm{eq}}$ are evenly generated between 0 km s$^{-1}$ 
and $V_{\mathrm{cri}}$. The high number fraction of fast rotators ($V\sin i >$ 200 
km s$^{-1}$) that we find in M11 may indicate that the underlying distribution of 
$V_{\mathrm{eq}}$ may be biased towards the high velocity regime (Supplementary 
Fig. 2). We therefore assumed a linear probability distribution following: 

\begin{equation}
P(V_{\mathrm{eq}}) = \alpha(V_{\mathrm{eq}}-396.18) + 1.
\end{equation}
\noindent where $\alpha$ is the slope of the distribution and a free parameter 
to be determined in Cases 3 and 4. The probability distribution is normalized 
at 396.18 km s$^{-1}$ which is the maximum $V_{\mathrm{cri}}$ in the Geneva 
isochrone adopting $\omega = 0.95$ and $\log t = 8.4$ [yr]$^{28}$. In addition, 
a distribution of spin axes uniformly oriented in a 3D space is seen to an observer 
as a distribution biased towards high $i$ because of projection effect$^{29}$. The 
probability distribution of $i$ in Cases 1 and 3 is then given by a uniform distribution 
in $\cos i$. We randomly selected $\cos i$ between 0 and 1 to consider the symmetric 
behaviour of the cosine function for $i$ larger than 90$^{\circ}$. $i$ were obtained 
from their inverse cosine values. 

Recently, a strong alignment of spin axes of stars was found in a few old 
open clusters$^{29,48}$. This fact allows us to consider a Gaussian probability 
distribution of $i$: 
\begin{equation}
P(i) = {1 \over \sqrt{ 2\pi\sigma_i^2}} e^{{(i - i_{\mathrm{peak}})^2 \over 2\sigma_i^2}}
\end{equation}
\noindent where the peak inclination angle ($i_{\mathrm{peak}}$) and the 
dispersion ($\sigma_i$) are free parameters to be determined in Cases 2 and 4.

We introduced the systematic and random errors estimated above 
to the $V\sin i$ of artificial stars. The simulated distributions of $V\sin i$ 
were then compared with the observed one using the K-S test. This simulation 
was repeated 1000 times to suppress the statistical fluctuation, and then the 
resultant probabilities were averaged. 

For Cases 2 -- 4, we tested $i_{\mathrm{peak}}$ in the range 5$^{\circ}$ to 
90$^{\circ}$ (with a step of 5$^{\circ}$), $\sigma _i$ in the range of 1$^{\circ}$ 
to 46$^{\circ}$ (with a step of 5$^{\circ}$), and $\alpha$ between 0 km$^{-1}$ s 
and 0.0023 km$^{-1}$ s (with a step of 0.0005 km$^{-1}$ s). 
The systematic and random errors were added to the resultant $V\sin i$. The 
distributions of the simulated $V\sin i$ for each case were then compared with the 
observed one using the K-S test. This provided the first estimation of each parameter. 
Refined simulations were then made around the first estimation using 20 per cent 
smaller steps. As a result, we found the best-fit parameters (Supplementary Figs. 8, 
9, and 10).\\

\noindent{\bf Gravity darkening effects.} We corrected for gravity darkening 
effects on the temperature and luminosity from the Roche potential model for 
rotating stars following the description of Ref. 49. A relation between the 
bolometric luminosity and the luminosity measured by an observer is expressed 
by Equation 6 of Ref. 49:

\begin{equation}
L(i) = L_{\mathrm{bol}} {4 \over \Sigma} \int_{d\Sigma \cdot d > 0} f^4(x,\omega,\theta) d\Sigma\cdot d = L_{\mathrm{bol}} C_{L}(i,\omega)
\end{equation}
\noindent where $\Sigma$, $d$, and $C_{L}(i,\omega)$ are the surface of 
a star and the direction of the observer inclined by $i$ from its spin axis, 
and the geometric correction term in luminosity, respectively. On the other 
hand, the effective temperature inferred by an observer can be obtained by 
averaging flux over the projected surface ($\Sigma_{\mathrm{p}}$) of a star. 
The relation between the mean effective temperature and the observed 
temperature is given by Equation 9 of Ref. 49 as below:

\begin{equation}
T_{\mathrm{eff}}(i,\omega) = T_{\mathrm{eff,mean}} \Bigg[{1 \over \Sigma_p} \int_{d\Sigma \cdot d > 0} f^4(x,\omega,\theta) d\Sigma\cdot d\Bigg]^{1\over4} = T_{\mathrm{eff,mean}} C_{T_{\mathrm{eff}}}(i,\omega)
\end{equation}
\noindent where $C_{T_{\mathrm{eff}}}(i,\omega)$ is the geometric correction 
term on the effective temperature. The geometric correction terms for luminosity and 
effective temperature were computed for 21 values of $i$ from 0$^{\circ}$ to 90$^{\circ}$ 
with an interval of 4.5$^{\circ}$ and for 21 values of $\omega$ ranging from 
0.0 to 1 with an interval of 0.05, respectively (Supplementary Fig. 3).\\

\noindent{\bf A synthetic colour-magnitude diagram.} We assigned masses, 
$V_{\mathrm{eq}}$, and $i$ to 1000 artificial stars based on the mass 
function$^{18}$ and the probability distributions from the Case 4 simulation. 
The other parameters of the artificial stars, such as effective temperature, 
bolometric luminosity, pole radius ($R_{\mathrm{p}}$), and oblateness, were 
obtained by interpolating their mass and $V_{\mathrm{eq}}$ to a grid of the 
Geneva isochrones ($\log t = 8.4$) in the $\omega$ range of 0 to 0.95$^{28}$.

The critical velocities of given artificial stars were computed by using the equation 
below:
\begin{equation}
V_{\mathrm{cri}} =\sqrt{{GM \over R_{\mathrm{eq}}}} = \sqrt{{2GM \over 3R_{\mathrm{p}}}}
\end{equation}
\noindent Their $\omega$ were computed according to the equation $\omega= 
{3 \over 2} {V_{\mathrm{eq}} \over V_{\mathrm{cri}}} {R_{\mathrm{p}} \over R_{\mathrm{eq}}}$. 
The gravity darkening correction values were obtained by interpolating the grids 
of the geometric correction values [$C_{L}(i,\omega)$ and $C_{T_{\mathrm{eff}}}(i,\omega$)] 
according to the values of $i$ and $\omega$ and then applied to the effective temperature 
and bolometric luminosity of the individual artificial stars, respectively.

The corrected effective temperature and luminosity were transformed to $U-V$ 
and $M_V$ using a colour-temperature relation and bolometric 
correction for the solar metallicity$^{50}$. A distance modulus ($V-M_V$) of 
11.55 mag and mean reddening $E(U-V)$ of 0.74 from a previous study$^{18}$ 
were applied to the synthetic colour-magnitude diagram. Non-negligible differential 
reddening was found in M11$^{18}$: $E(B-V)$ ranges from 0.38 to 0.48 mag with a standard 
deviation of 0.03 mag, which corresponds to 0.05 mag in $E(U-V)$. The differential 
reddening was randomly introduced to the synthetic colour-magnitude diagram 
assuming a Gaussian distribution with the standard deviation. The photometric 
errors in $U-V$ and $V$ provided from the published data$^{18}$ 
were averaged within each $V$ magnitude bin of 1 mag. These mean errors were 
adopted as the dispersions of the Gaussian distributions. Photometric errors generated 
from these distributions were added to the synthetic colour-magnitude 
diagram. About 50 stars were brightened by 0.75 mag adopting the minimum binary fraction 
of 5 per cent. We found that the adopted $U-V$ colour scale$^{50}$ appears to be, on 
average, about 0.05 bluer than the one$^{51}$ used in the reddening determination of 
the previous study$^{18}$ (see Supplementary Fig. 11). The $U-V$ colours of the artificial 
stars were thus shifted by such a systematic difference. \\

\noindent{\bf Data Availability.} In this paper, we use publicly available data : photometry from Ref. 18, evolutionary tracks from the Geneva stellar evolution group (Ref. 28, \url{https://www.unige.ch/sciences/astro/evolution/en/?lang=en}), astrometry from {\it Gaia} DR2 (\url{https://www.cosmos.esa.int/gaia}), and spectra from {\it Gaia}-ESO survey (\url{https://www.gaia-eso.eu/data-products/public-data-releases}). New MMT H$\alpha$ spectra (shown in Supplementary Fig. 1) and derived $V\sin i$ (used in Figs 2, 3, 4 , and Supplementary. Fig. 2) are available for download at \url{ftp://ftp.astro.ulg.ac.be/pub/users/lim} (anonymous ftp).\\

\noindent{\bf References}\\
31. McNamara, B. J., Pratt, N. M., \& Sanders, W. L. Membership in the open cluster M11, {\it Astron. Astrophys. Suppl.} {\bf 27,} 117-143 (1977)\\
32. Su, C.-G., Zhao, J.-L., \& Tian, K.-P. Membership determination of stars using proper motions in the region of the open cluster M 11, {\it Astron. Astrophys. Suppl.} {\bf 128,} 255-264 (1998)\\
33. Gaia Collaboration {\it et al.} The Gaia mission. {\it Astron. Astrophys.} {\bf 595,} 1-36 (2016)\\
34. Gaia Collaboration {\it et al.} Gaia Data Release 2. Summary of the contents and survey properties. {\it eprint arXiv:1804.09365} (2018)\\
35. Szentgyorgyi, A. {\it et al.} Hectochelle: A Multiobject Optical Echelle Spectrograph for the MMT, {\it Publ. Astron. Soc. Pac.} {\bf 123}, 1188-1209 (2011)\\
36. Lim, B. {\it et al.} A Constraint on the Formation Timescale of the Young Open Cluster NGC 2264: Lithium Abundance of Pre-main Sequence Stars. {\it Astrophys. J.} {\bf 831,} 116-133 (2016)\\
37. Lim, B. {\it et al.} Kinematic evidence for feedback-driven star formation in NGC 1893. {\it Mon. Not. R. Astron. Soc.} {\bf 477,} 1993-2003 (2018)\\
38. Pasquini, L. {\it et al.} Installation and commissioning of FLAMES, the VLT Multifibre Facility. {\it The Messenger} {\bf 110}, 1-9 (2002)\\
39. Dekker, H. {\it et al.} Design, construction, and performance of UVES, the echelle spectrograph for the UT2 Kueyen Telescope at the ESO Paranal Observatory. in {\it SPIE Proceedings: Optical and IR Telescope Instrumentation and Detectors}. 534-545 (2000)\\
40. Merle, T. {\it et al.} The Gaia-ESO Survey: double-, triple-, and quadruple-line spectroscopic binary candidates, {\it Astron. Astrophys.} {\bf 608}, 95-128 (2017)\\
41. Koo, J.-R. {\it et al.} Variable Stars in the Open Cluster M11 (NGC 6705). {\it Publ. Astron. Soc. Pac.} {\bf 119,} 1233-1246 (2007)\\
42. Brown, A. G. A., \& Verschueren, W. High S/N Echelle spectroscopy in young stellar groups II. Rotational velocities of early-type stars in Sco OB2, {\it Astron. Astrophys.} {\bf 319}, 811-838 (1997)\\
43. Royer, F., Zorec, J., \& G\'omez, A. E. Rotational velocities of A-type stars III. Velocity distributions, {\it Astron. Astrophys.} {\bf 463}, 671-582 (2007)\\
44. Royer, F., Gerbaldi, M., Faraggiana, R., \& G\'omez, A. E. Rotational velocities of A-type stars I. Measurement of $v\sin i$ in the southern hemisphere, {\it Astron. Astrophys.} {\bf 381}, 105-121 (2002)\\
45. Al-Naimiy, H. M. Linearized limb-darkening coefficients for use in analysis of eclipsing binary light curves. {\it Astrophys. Sp. Sci.} {\bf 53,} 181-192 (1978)\\ 
46. Castelli, F. \& Kurucz, R. L. New Grids of ATLAS9 Model Atmospheres. {\it eprint arXiv:astro-ph/0405087} (2004).\\
47. Gray, D. Precise Rotation Rates for Five Slowly Rotating a Stars. {\it Astron. J.} {\bf 147,} 81-93 (2014)\\
48. Kovacs, G. Signature of non-isotropic distribution of stellar rotation inclination angles in the Praesepe cluster. {\it Astron. Astrophys.} {\bf 612,} 2-6 (2018)\\
49. Georgy, C. {\it et al.} Populations of rotating stars III. SYCLIST, the new Geneva population synthesis code. {\it Astron. Astrophys.} {\bf 566,} 21-35 (2014)\\
50. Worthey, G. \& Lee, H.-C. An Empirical UBV RI JHK Color-Temperature Calibration for Stars. {\it Astrophys. J. Suppl. Ser.} {\bf 193,} 1-11 (2011)\\
51. Mermilliod, J.-C. Comparative studies of young open clusters. III - Empirical isochronous curves and the zero age main sequence. {\it Astron. Astrophys.} {\bf 97,} 235-244 (1981)\\

\noindent {\bf Acknowledgements}\\ 
The authors thank Michael Bessell, Sylvia Ekstr\"om and David Gray 
for valuable comments and also thank Perry Berlind, Mike Calkins, Chun Ly, 
ShiAnne Kattner, and Nelson Caldwell for assisting with Hectochelle observations. 
BL is grateful for Seulgi Kim's assistance in running the 
simulation codes. This paper has used the data obtained under the K-GMT 
Science Program (PID: MMT-2017A-1) funded through Korean GMT Project operated 
by KASI). This work has made use of data from the European Space Agency (ESA) mission
{\it Gaia} (\url{https://www.cosmos.esa.int/gaia}), processed by the {\it Gaia}
Data Processing and Analysis Consortium (DPAC,
\url{https://www.cosmos.esa.int/web/gaia/dpac/consortium}). Funding for the DPAC
has been provided by national institutions, in particular the institutions
participating in the {\it Gaia} Multilateral Agreement. B.L. and H.S. acknowledge 
the support of the National Research Foundation of Korea, Grant No. NRF-2017R1A6A3A03006413 
and NRF-2015R1D1A1A01058444, respectively. YN and GR also acknowledge the 
support by the FNRS and the PRODEX contract, respectively.\\

\noindent {\bf Author Contributions}\\
BL proposed this project, analyzed the data, and wrote the manuscript. 
GR computed the gravity darkening effects. YN developed the Fourier 
transform code. HS was involved in the planning the project. All co-authors 
participated in discussion and contributed to improvement of the manuscript. \\

\noindent {\bf Competing Interests}\\
The authors declare that they have no competing financial interests.\\

\noindent {\bf Correspondence}\\
Correspondence and requests for materials should be addressed to Beomdu Lim 
(email:blim@uliege.be).\\

\newpage

\begin{figure}[!t]
\centering
\setcounter{figure}{0}
\renewcommand{\figurename}{Supplementary Figure}
\includegraphics[angle=0,width=150mm]{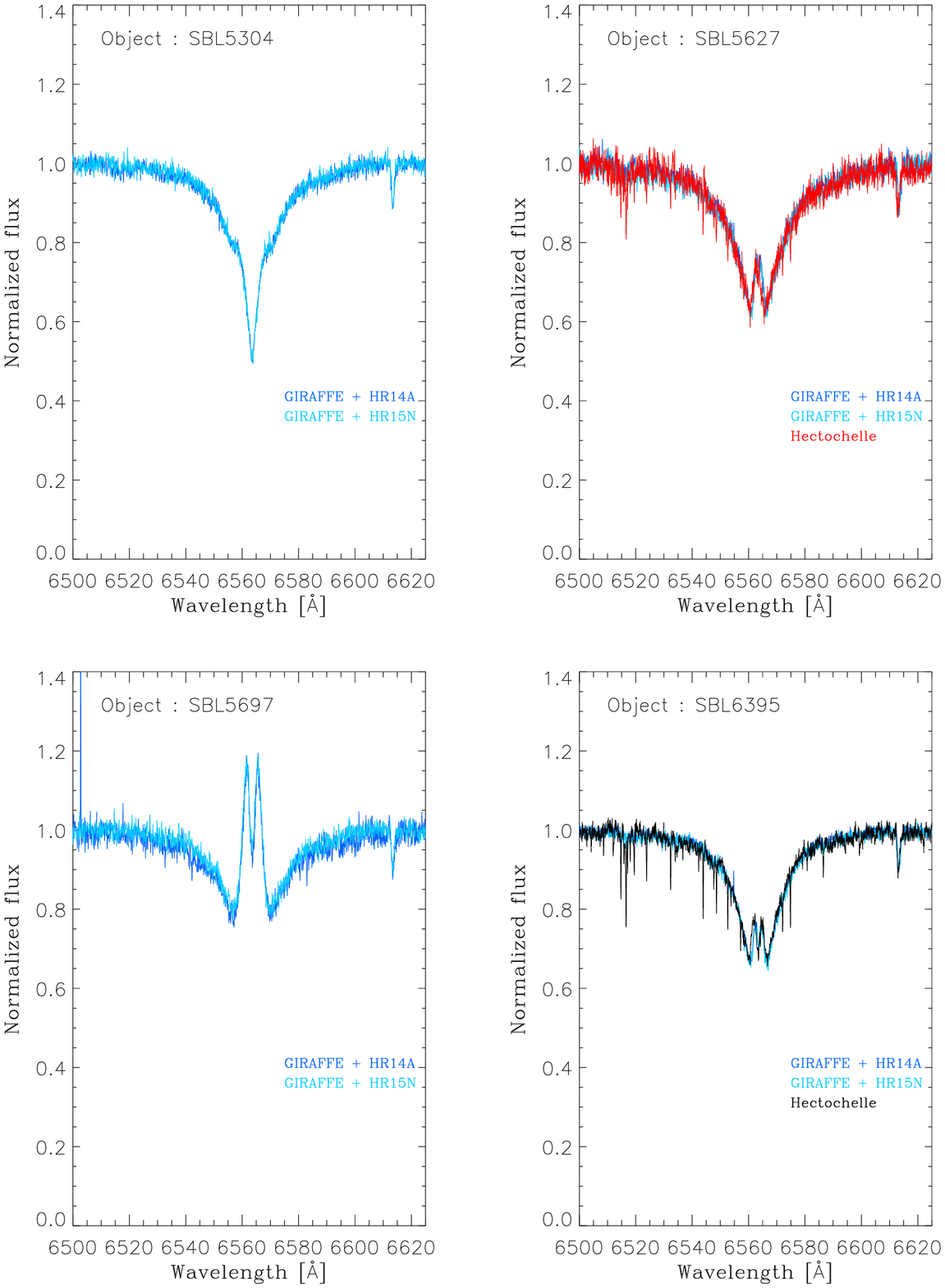}
\caption{{\bf H$\alpha$ emission line of Be stars.} These spectra were taken 
from three different spectrographs FLAMES/GIRAFFE (blue and cyan), UVES (green), 
and Hectochelle (black, yellow green, and red).}
\vspace{0mm} %% add extra space ONLY when figures/tables are "colliding"!
\end{figure}
\clearpage

\begin{figure}[!t]
\centering
\setcounter{figure}{0}
\renewcommand{\figurename}{Supplementary Figure}
\includegraphics[angle=0,width=150mm]{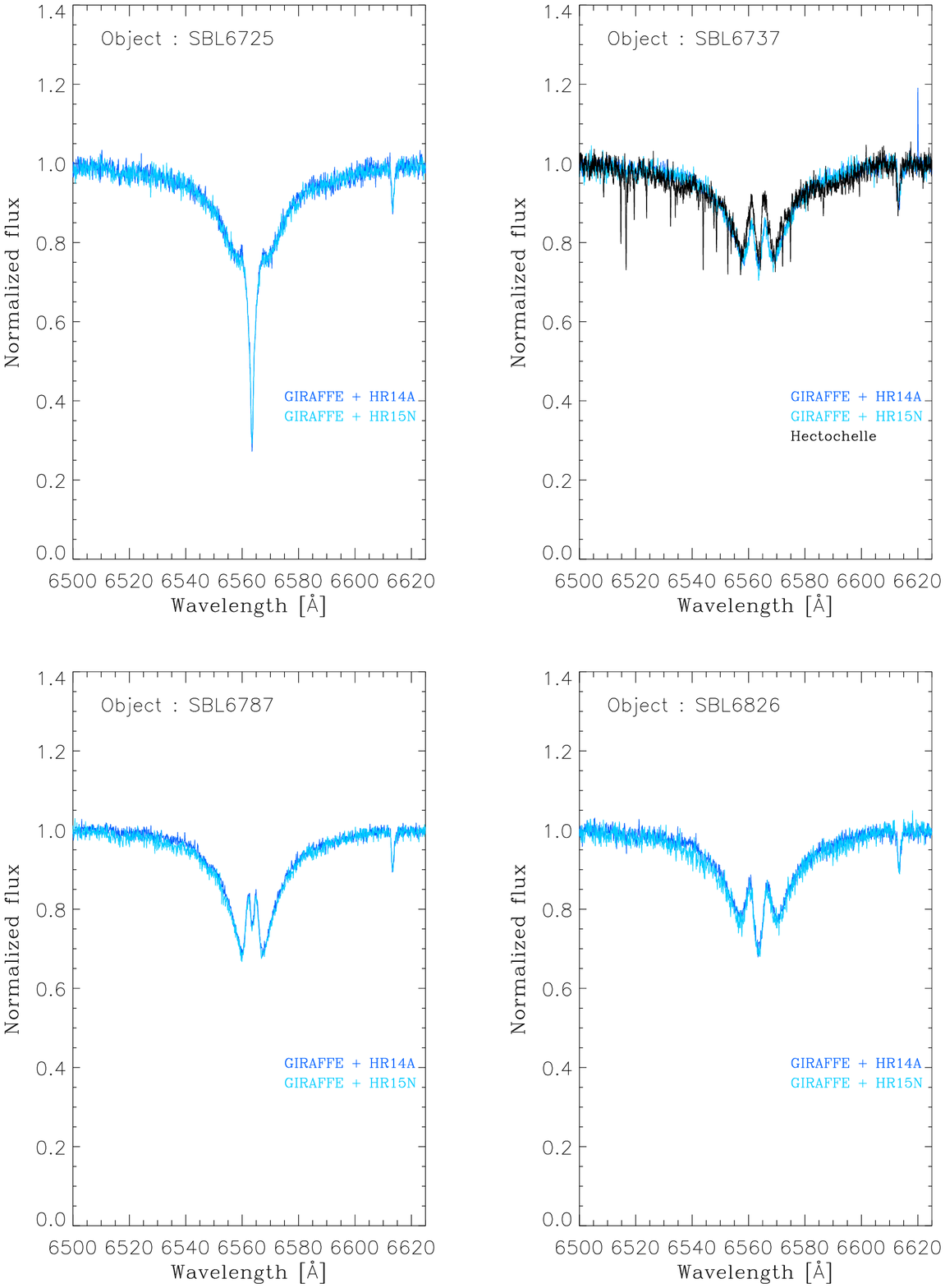}
\caption{(Continue)}
\vspace{0mm} %% add extra space ONLY when figures/tables are "colliding"!
\end{figure}
\clearpage

\begin{figure}[!t]
\centering
\setcounter{figure}{0}
\renewcommand{\figurename}{Supplementary Figure}
\includegraphics[angle=0,width=150mm]{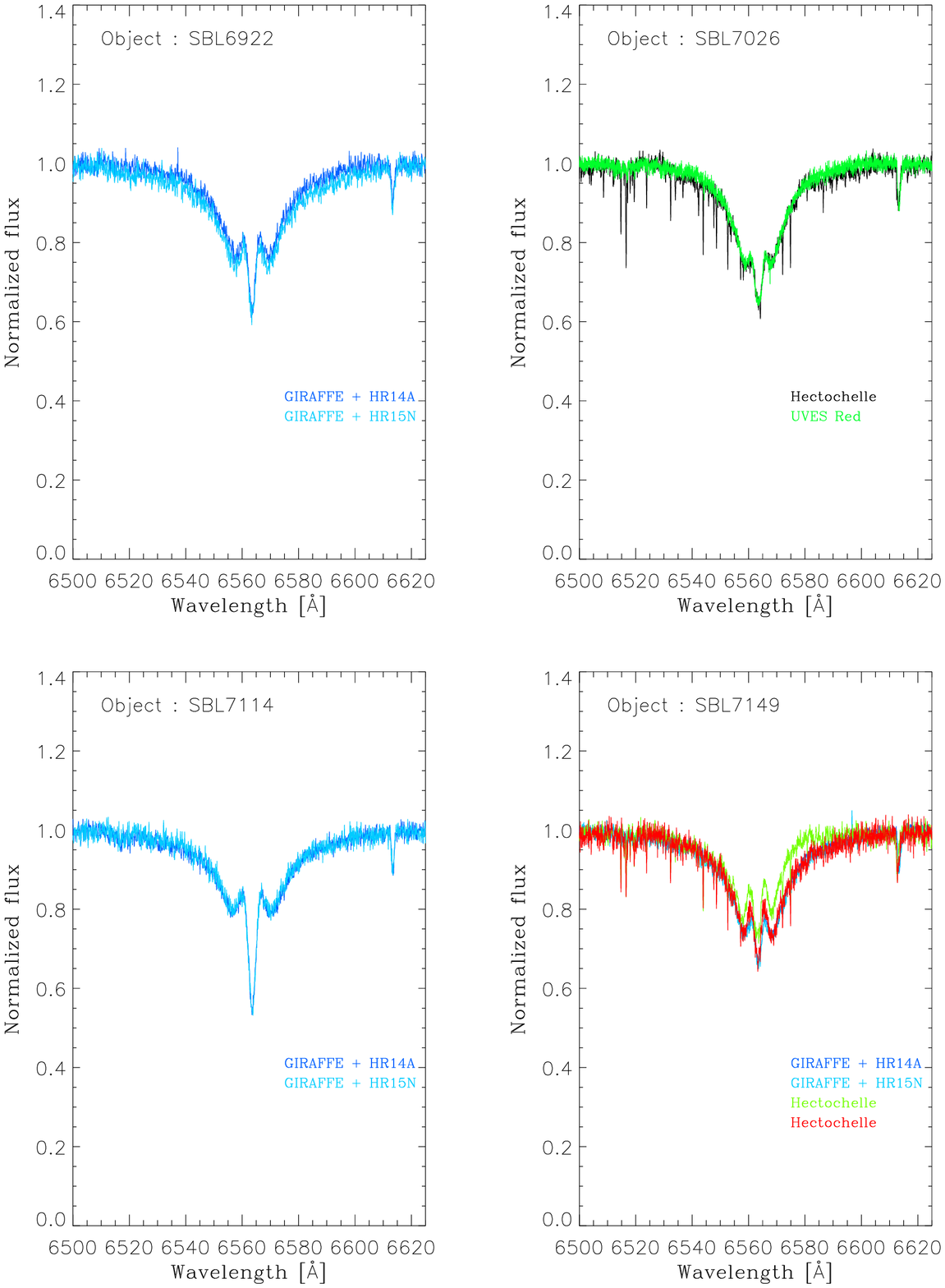}
\caption{(Continue)}
\vspace{0mm} %% add extra space ONLY when figures/tables are "colliding"!
\end{figure}
\clearpage

\begin{figure}[!t]
\centering
\setcounter{figure}{0}
\renewcommand{\figurename}{Supplementary Figure}
\includegraphics[angle=0,width=150mm]{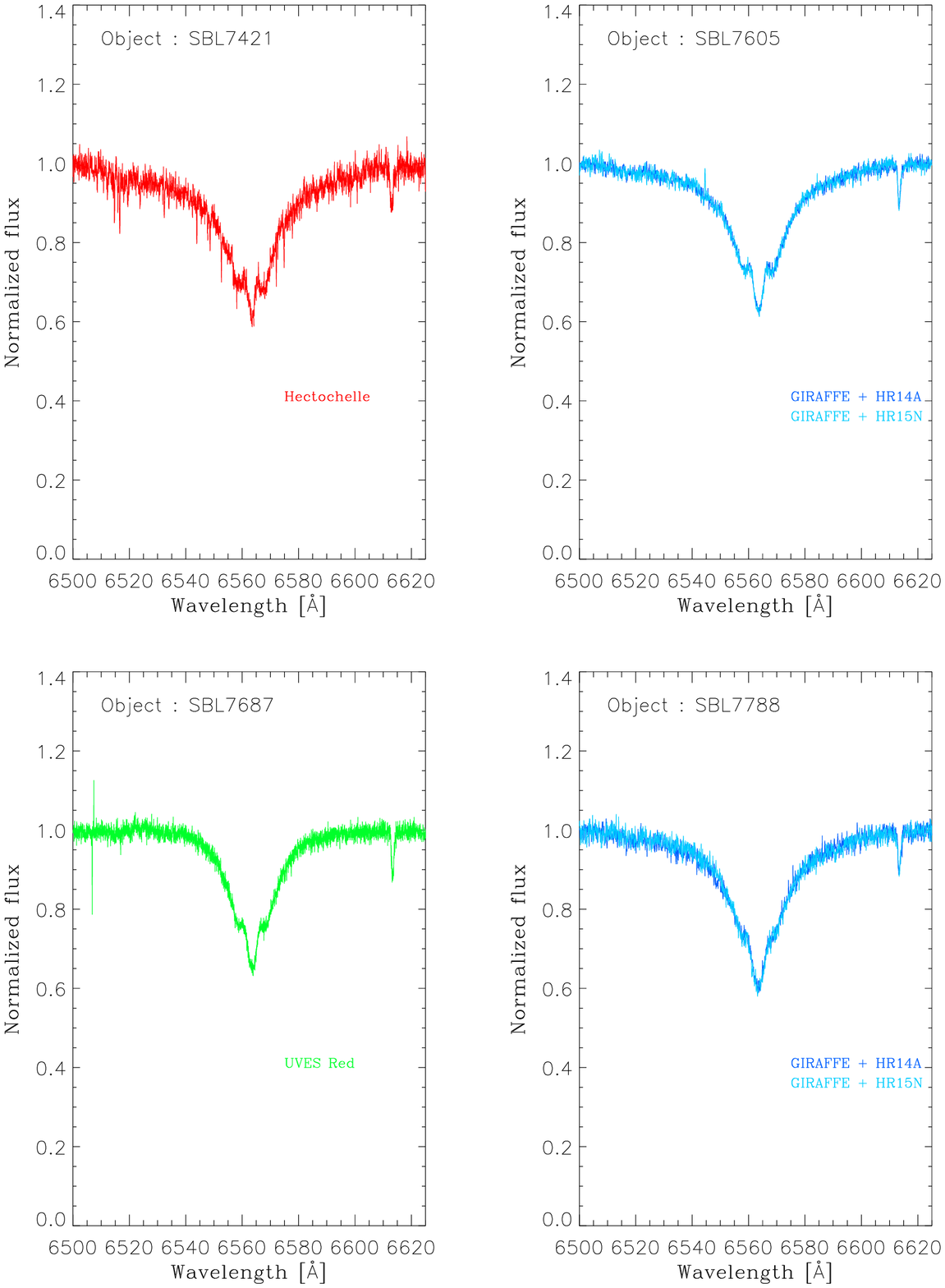}
\caption{(Continue)}
\vspace{0mm} %% add extra space ONLY when figures/tables are "colliding"!
\end{figure}
\clearpage

\begin{figure}[!t]
\centering
\setcounter{figure}{0}
\renewcommand{\figurename}{Supplementary Figure}
\includegraphics[angle=0,width=150mm]{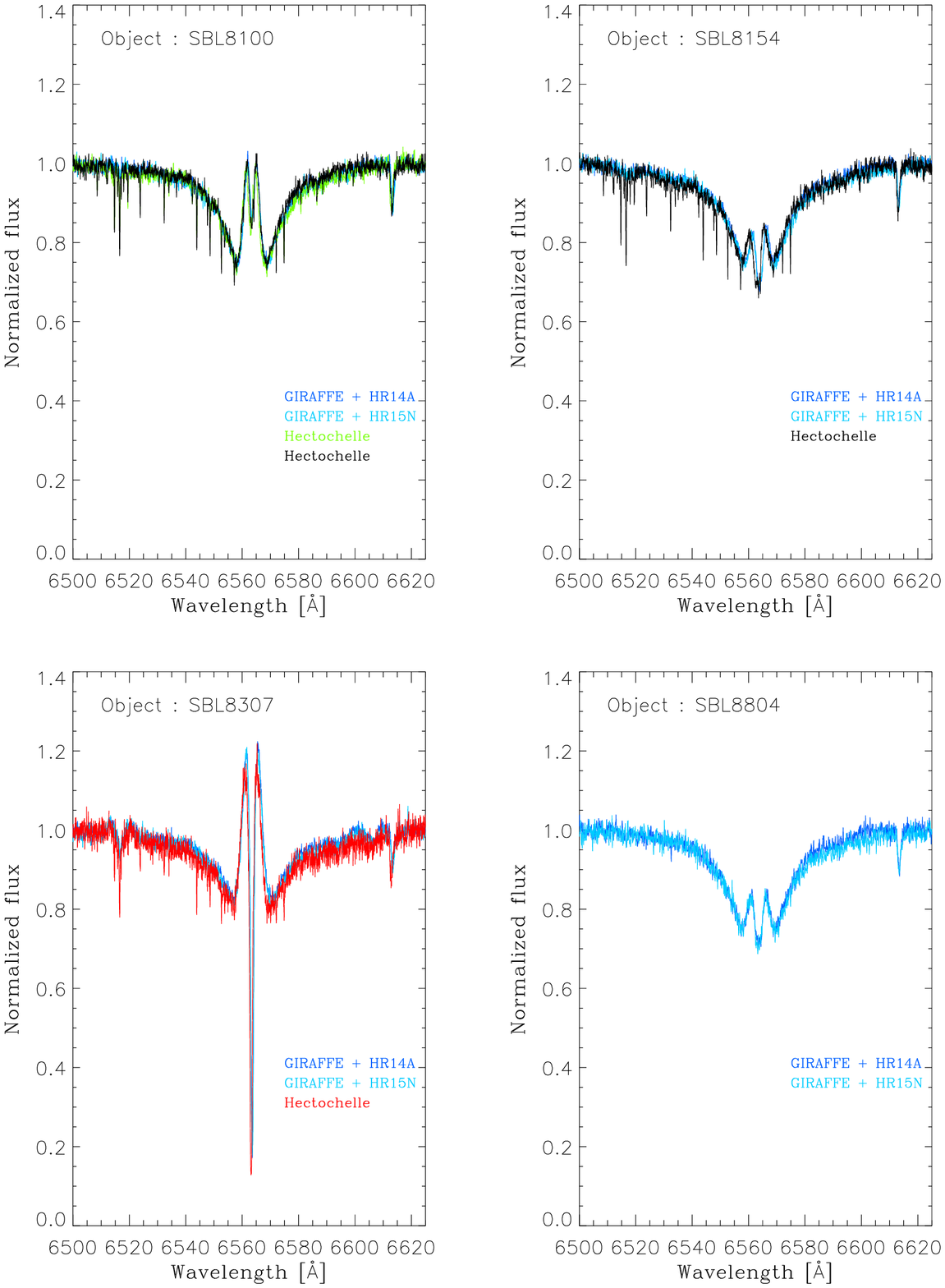}
\caption{(Continue) }
\vspace{0mm} %% add extra space ONLY when figures/tables are "colliding"!
\end{figure}
\clearpage

\begin{figure}[!t]
\centering
\setcounter{figure}{0}
\renewcommand{\figurename}{Supplementary Figure}
\includegraphics[angle=0,width=150mm]{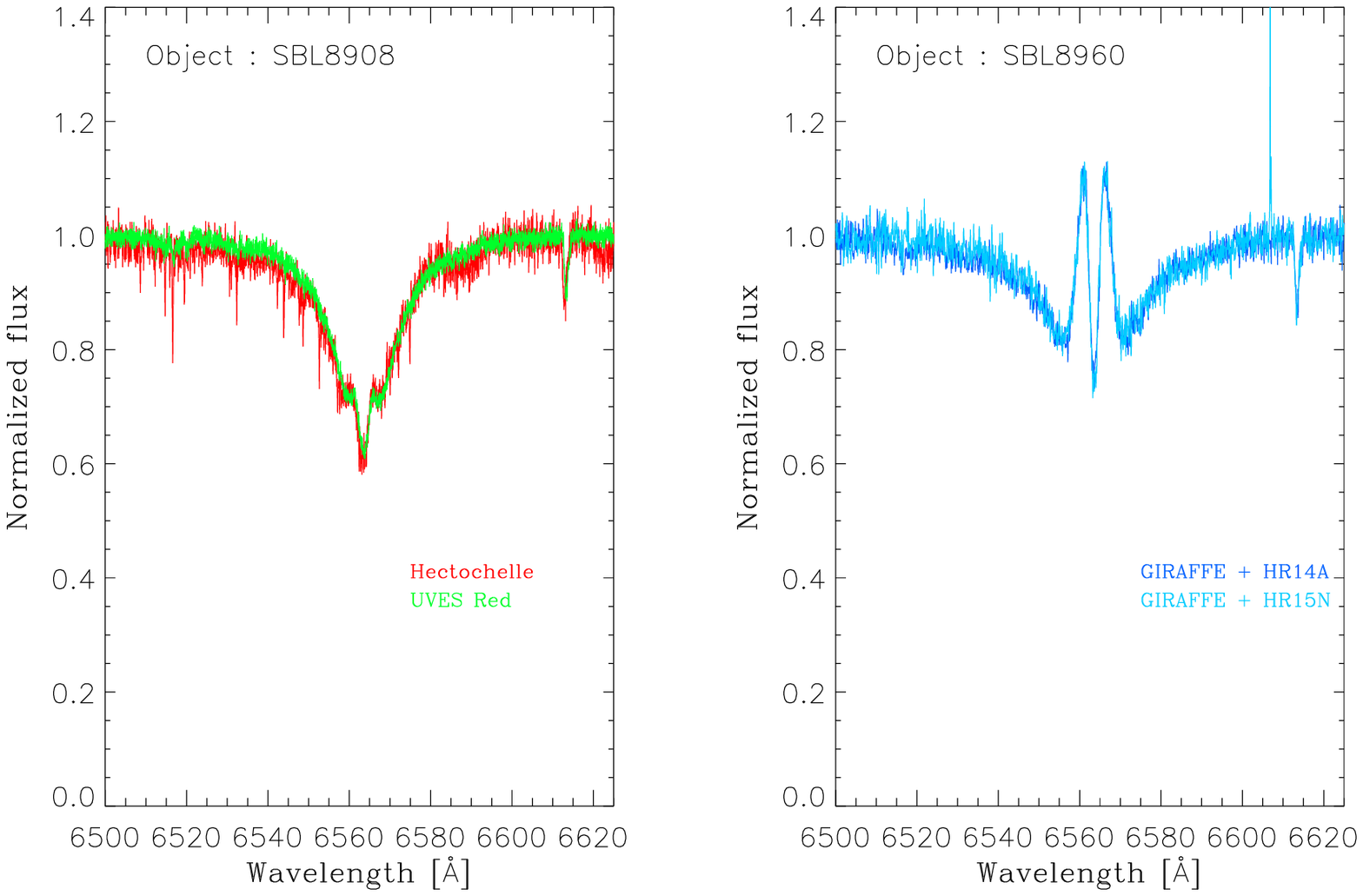}
\caption{(Continue)}
\vspace{0mm} %% add extra space ONLY when figures/tables are "colliding"!
\end{figure}
\clearpage

\begin{figure}[!t]
\centering
\setcounter{figure}{1}
\renewcommand{\figurename}{Supplementary Figure}
\includegraphics[angle=0,width=150mm]{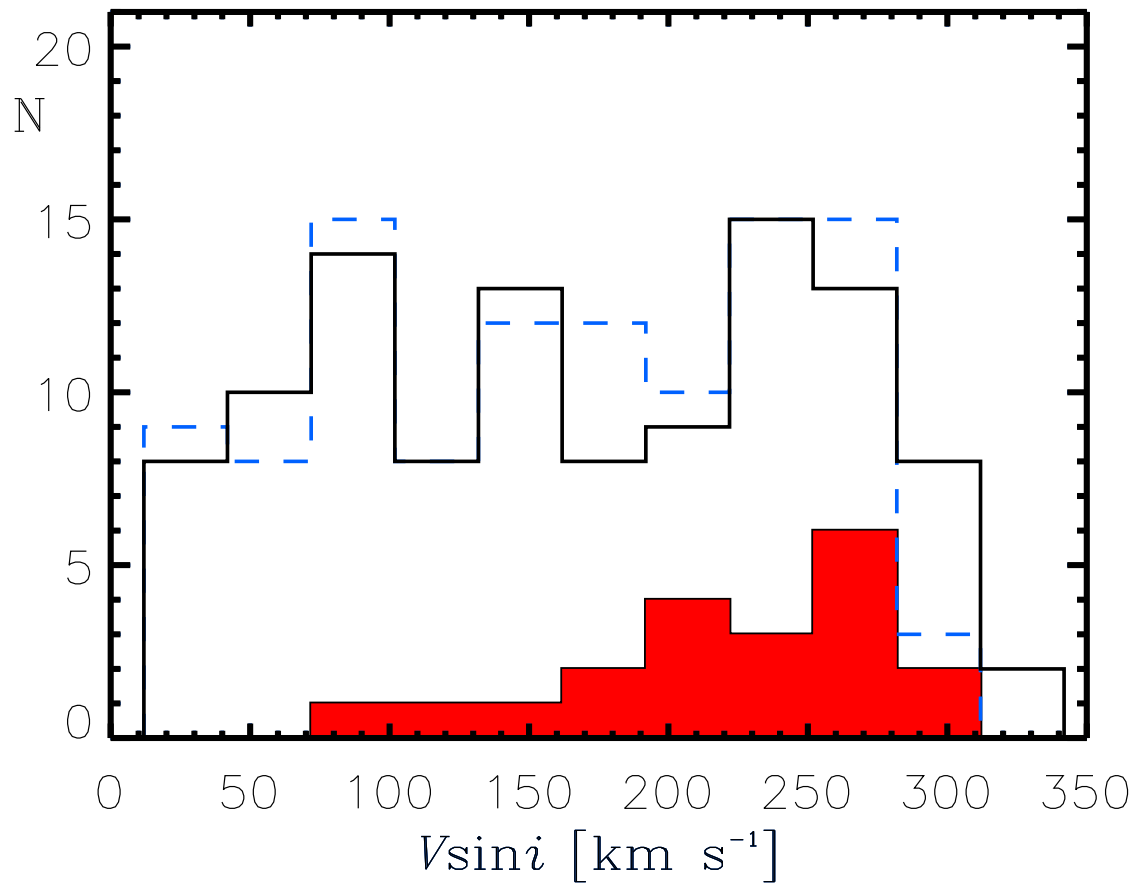}
\caption{{\bf Distribution of $V\sin i$.} Solid and dashed line histograms 
represent the distributions of the observed $V\sin i$ and the systematic 
error-corrected $V\sin i$, respectively. The red histogram shows the $V\sin i$ 
distribution of Be stars. All the histograms were obtained with a bin size of 
30 km s$^{-1}$.}
\vspace{0mm} %% add extra space ONLY when figures/tables are "colliding"!
\end{figure}
\clearpage

\begin{figure}[!t]
\centering
\setcounter{figure}{2}
\renewcommand{\figurename}{Supplementary Figure}
\includegraphics[angle=0,width=160mm]{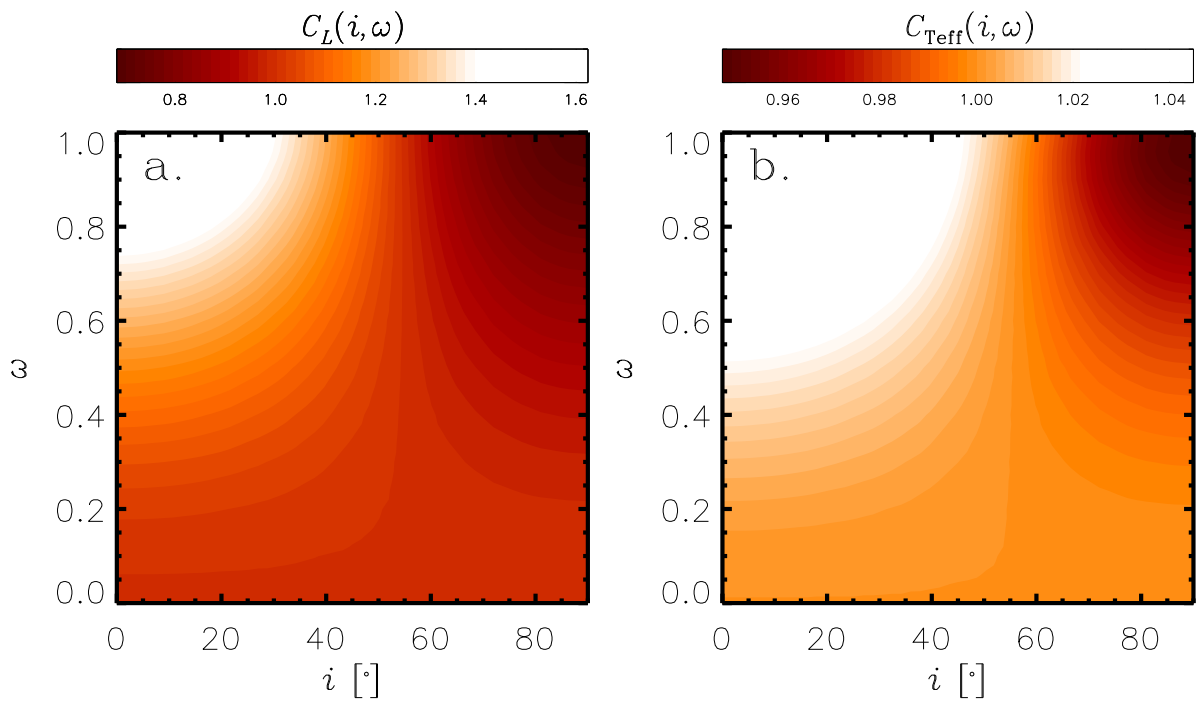}
\caption{{\bf Gravity darkening effects on luminosity and effective temperature.} 
{\bf a.} The geometric correction term for luminosity with respect to $i$ and 
$\omega$. {\bf b.} The geometric correction term for effective temperature. The 
correction values were plotted on a linear scale.  }
\vspace{0mm} %% add extra space ONLY when figures/tables are "colliding"!
\end{figure}
\clearpage

%\begin{figure}[!t]
%\centering
%\setcounter{figure}{3}
%\renewcommand{\figurename}{Supplementary Figure}
%\includegraphics[angle=0,width=160mm]{sfig4.eps}
%%\setlength{\abovecaptionskip}{-3pt}
%\caption{{\bf Gravity darkening effects on colour-magnitude diagrams 
%with various inclination angle distributions.} Different $i_{\mathrm{peak}}$ 
%(20$^{\circ}$, 50$^{\circ}$, and 80$^{\circ}$) were applied to the diagrams, 
%respectively, while the other parameters are the same as those derived from the 
%Case 4 simulations. A colour-magnitude diagram adopting a random orientation of $i$ is plotted 
%in the lower-right side panel for comparison.}
%\vspace{0mm} %% add extra space ONLY when figures/tables are "colliding"!
%\end{figure}
%\clearpage

\begin{figure}[!t]
\centering
\setcounter{figure}{3}
\renewcommand{\figurename}{Supplementary Figure}
\includegraphics[angle=0,width=160mm]{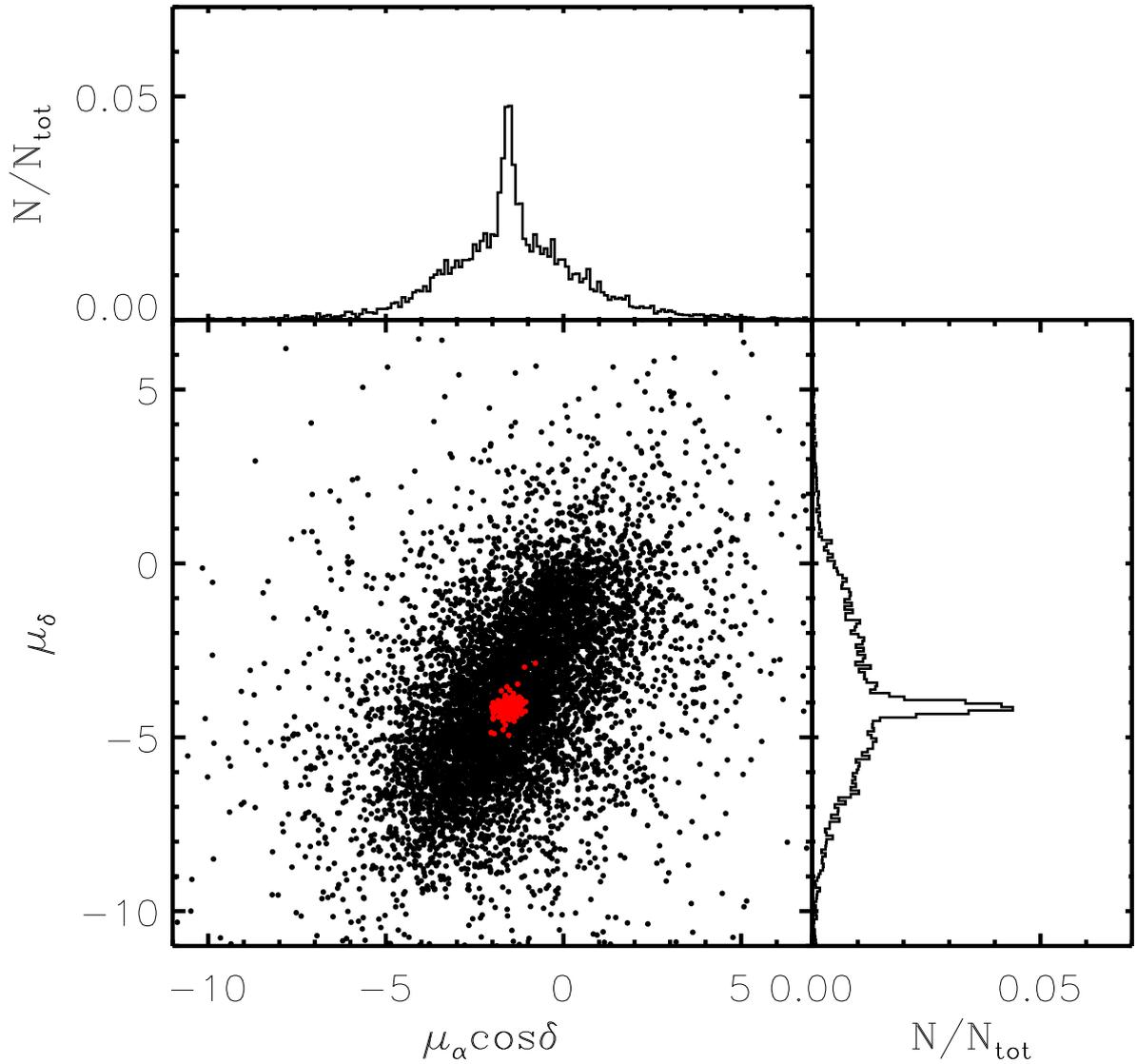}
\caption{{\bf Distribution of proper motion of stars from Gaia Data Release 2$^{33,34}$.} 
Black dots are all the stars observed by Gaia in the M11 field, while red dots are 
the members selected from previous proper motion data$^{31,32}$. Histograms 
were obtained along each proper motion direction. }
\vspace{0mm} %% add extra space ONLY when figures/tables are "colliding"!
\end{figure}
\clearpage

\begin{figure}[!t]
\centering
\setcounter{figure}{4}
\renewcommand{\figurename}{Supplementary Figure}
\includegraphics[angle=0,width=150mm]{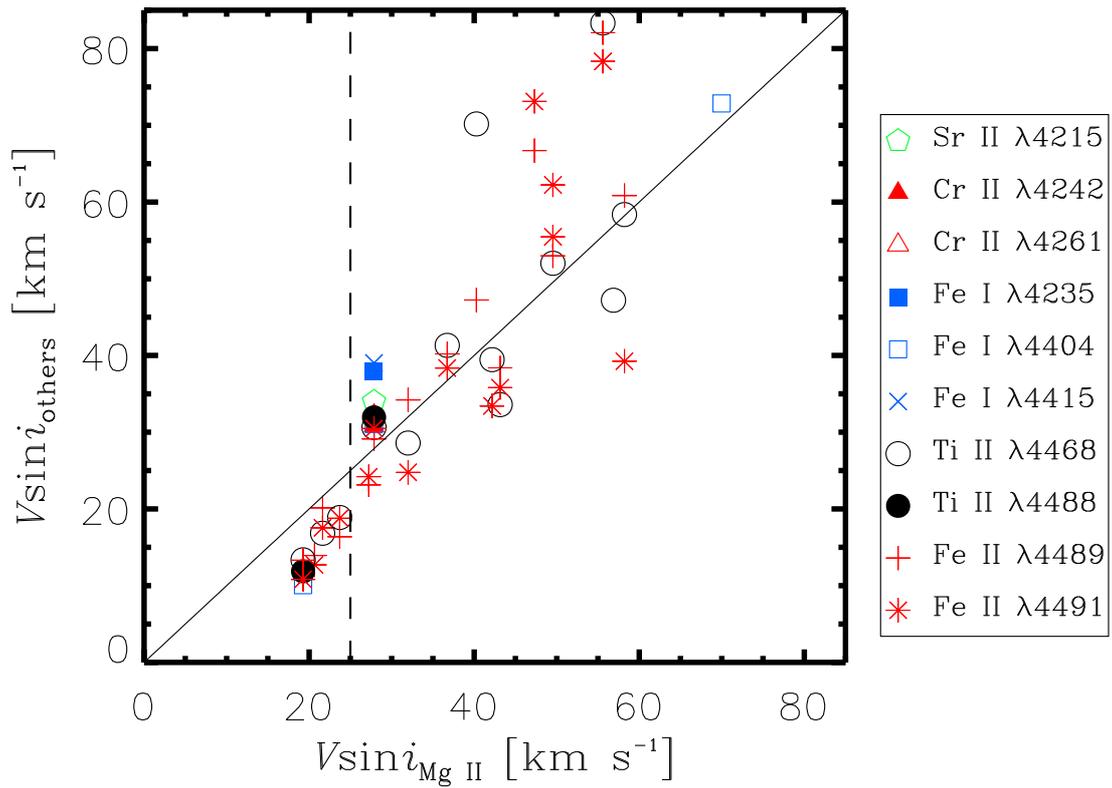}
\caption{{\bf Comparison of $V\sin i$ derived from Mg {\scriptsize \textsc{II}} 
$\lambda 4481$ with those from the other metallic lines.} Solid line represents 
the one-to-one correspondence of two measurements. Dashed line denotes 
the lower limit of measurable $V\sin i$ from the Mg {\scriptsize \textsc{II}} 
$\lambda 4481$ line. }
\vspace{0mm} %% add extra space ONLY when figures/tables are "colliding"!
\end{figure}
\clearpage

\begin{figure}[!t]
\centering
\setcounter{figure}{5}
\renewcommand{\figurename}{Supplementary Figure}
\includegraphics[angle=0,width=160mm]{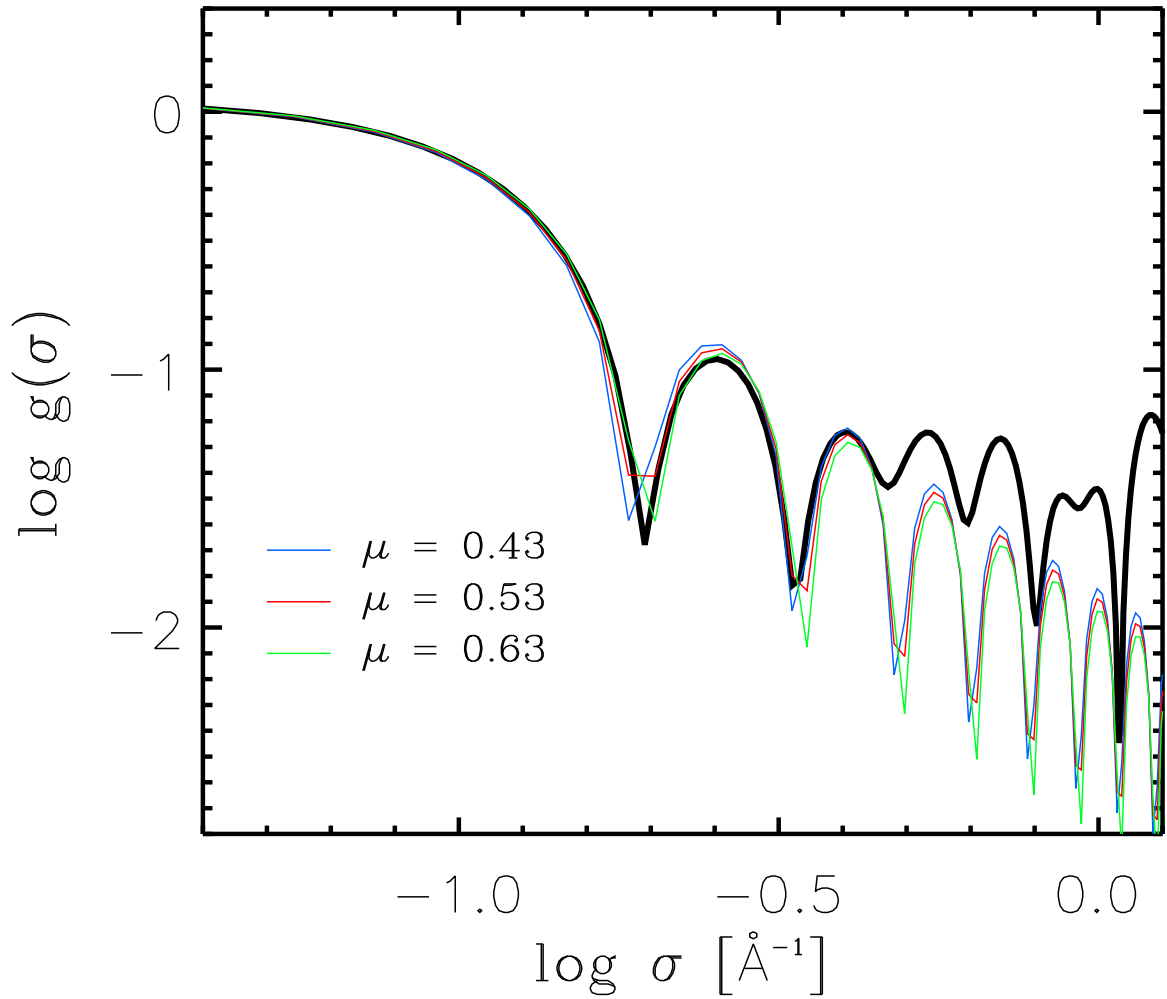}
\caption{{\bf Fourier spectrum of the Mg {\scriptsize \textsc{II}} 
$\lambda 4481$ line for a star in our sample (thick solid line).} The thin solid line represents 
the Fourier spectra of a pure rotation profile adopting 
different limb darkening coefficients, respectively.}
\vspace{0mm} %% add extra space ONLY when figures/tables are "colliding"!
\end{figure}
\clearpage

\begin{figure}[!t]
\centering
\setcounter{figure}{6}
\renewcommand{\figurename}{Supplementary Figure}
\includegraphics[angle=0,width=72.5mm]{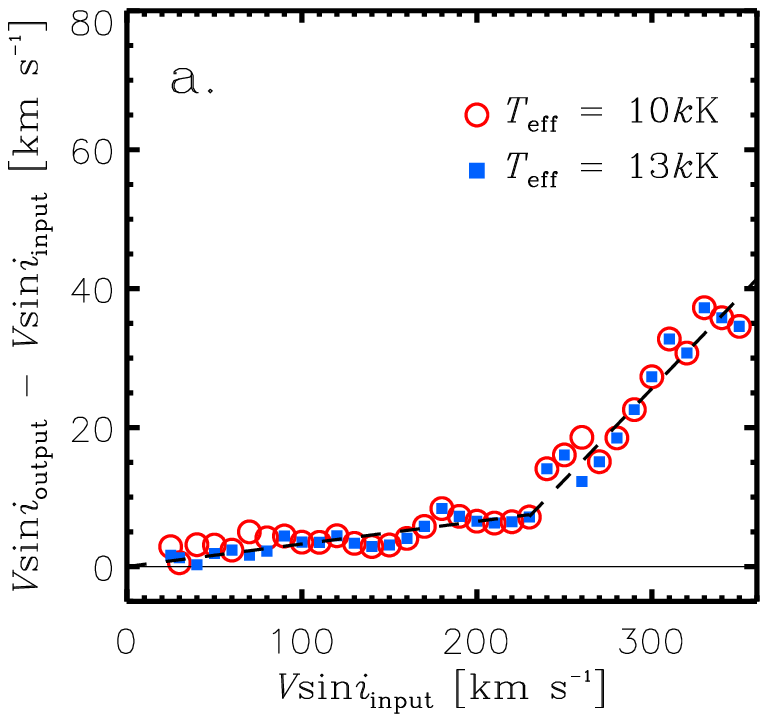}\includegraphics[angle=0,width=80mm]{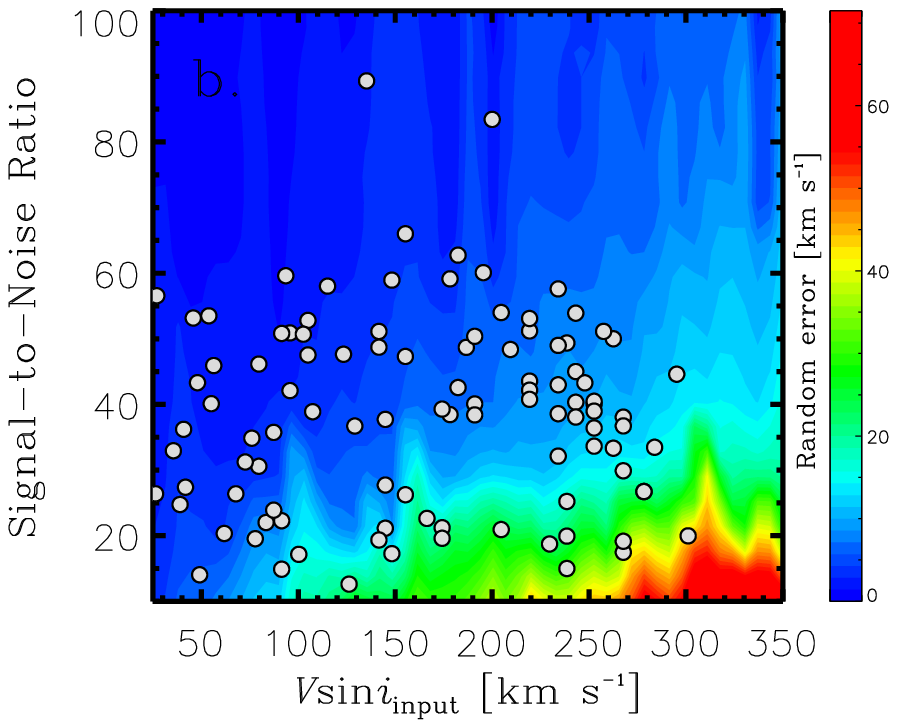}
\caption{{\bf Estimation of measurement errors.} {\bf a.} Systematic errors as a function of 
input $V\sin i$. The systematic errors involved in $V\sin i$ are not affected by effective 
temperature. Dashed lines show the results from linear least-square fitting to the systematic 
difference. {\bf b.} Distribution of random errors with respect to input $V\sin i$ and 
SNR. Dots in the right panel denote the systematic error-corrected $V\sin i$ of 
observed stars and the SNRs of their spectra.}
\vspace{0mm} %% add extra space ONLY when figures/tables are "colliding"!
\end{figure}
\clearpage

\begin{figure}[!t]
\centering
\setcounter{figure}{7}
\renewcommand{\figurename}{Supplementary Figure}
\includegraphics[angle=0,width=150mm]{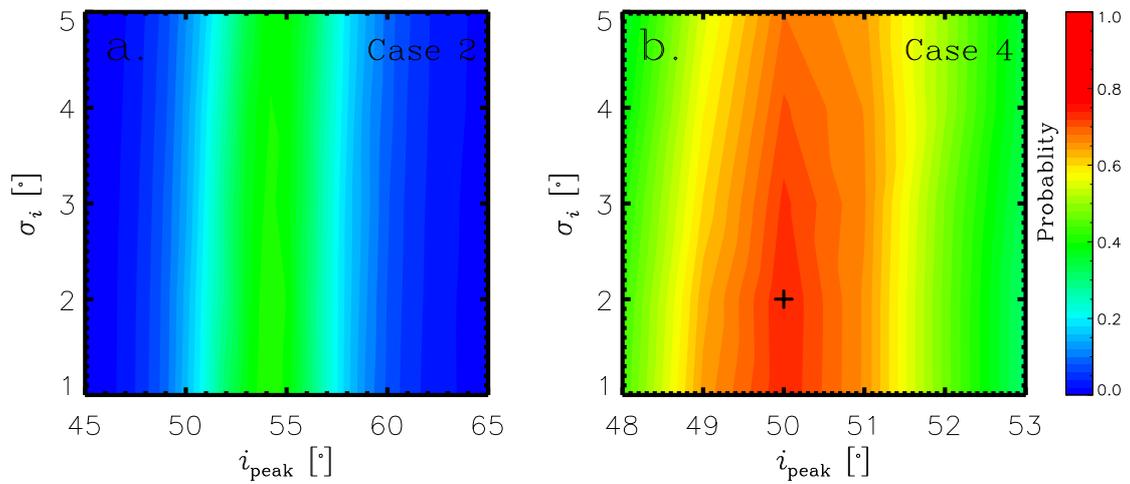}
\caption{{\bf Probability distributions with respect to $i_{\mathrm{peak}}$ and 
$\sigma_i$.} The contours shows the distribution of mean probabilities obtained 
from 1000 times simulations for each case, where the probabilities were computed 
with the K-S test between $V\sin i$ distributions for simulations and observations. 
{\bf a.} Parameter space for the Case 2 simulations. The mean probabilities appears 
high between 50$^{\circ}$ and 53$^{\circ}$. However, a specific value of $\sigma _i$ 
cannot be well constrained because the mean probabilities are almost the same 
in the range of 1$^{\circ}$ to 5$^{\circ}$. {\bf b.} Parameter 
space for the Case 4 simulation. The cross indicates $i_{\mathrm{peak}}$ and 
$\sigma_i$ showing highest probability with which to properly reproduce the 
observed $V\sin i$.  }
\vspace{0mm} %% add extra space ONLY when figures/tables are "colliding"!
\end{figure}
\clearpage

\begin{figure}[!t]
\centering
\setcounter{figure}{8}
\renewcommand{\figurename}{Supplementary Figure}
\includegraphics[angle=0,width=80mm]{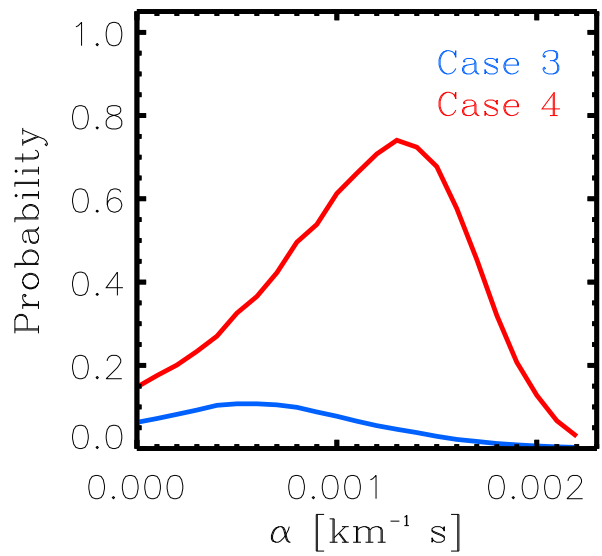}
\caption{{\bf Probability distributions with respect to $\alpha$.} Blue and red 
solid lines represent the variation of the mean probabilities from the K-S test for 
the Cases 3 and 4 simulations, respectively.}
\vspace{0mm} %% add extra space ONLY when figures/tables are "colliding"!
\end{figure}
\clearpage

\begin{figure}[!t]
\centering
\setcounter{figure}{9}
\renewcommand{\figurename}{Supplementary Figure}
\includegraphics[angle=0,width=160mm]{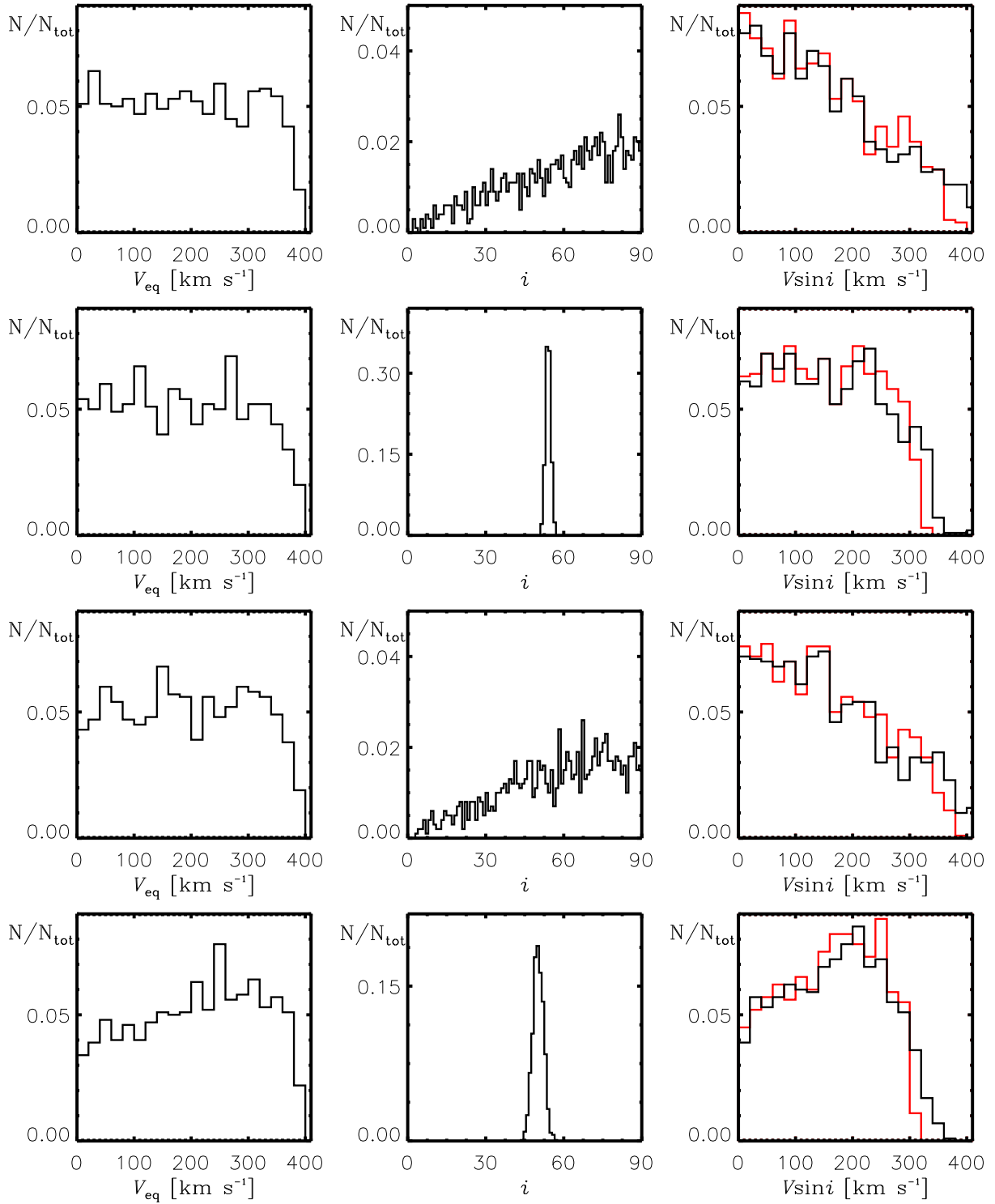}
\caption{{\bf Examples of underlying distributions inferred from each simulation.} 
Left-hand, middle, and right-hand panels show the distributions 
of $V_{\mathrm{eq}}$, $i$, and $V\sin i$, respectively. In the right-hand panels, red 
histograms represent the distributions of the simulated $V\sin i$, while 
black histograms display those of the error-added $V\sin i$. The 
examples generated for Cases 1 to 4 are plotted from top to bottom, respectively. }
\vspace{0mm} %% add extra space ONLY when figures/tables are "colliding"!
\end{figure}
\clearpage

\begin{figure}[!t]
\centering
\setcounter{figure}{10}
\renewcommand{\figurename}{Supplementary Figure}
\includegraphics[angle=0,width=100mm]{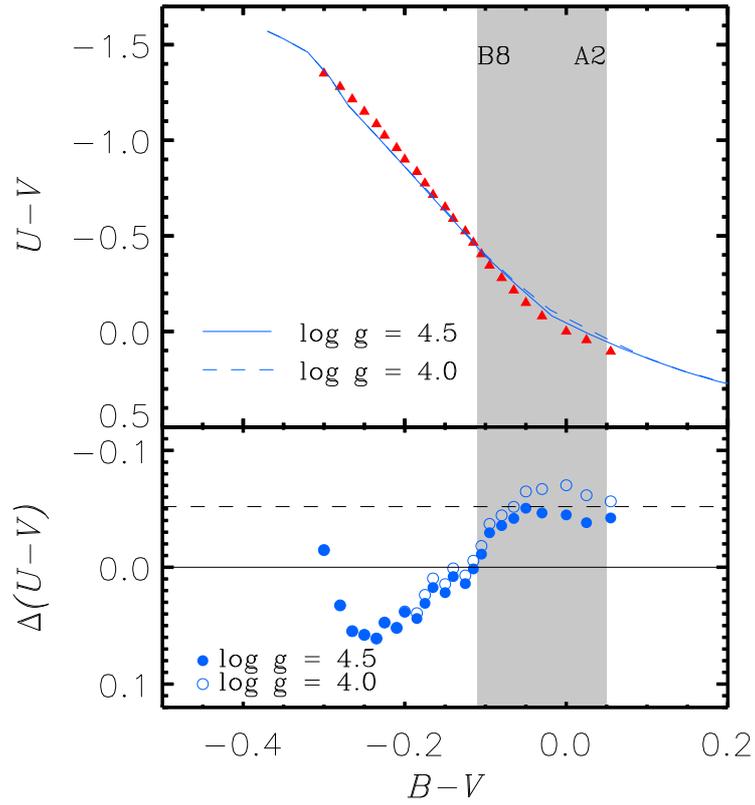}
\caption{{\bf Systematic difference between two different colour scales in $U-V$.} 
In the top panel, triangle denotes the $(U-V, B-V)$ relation of Ref. 51. Solid and 
dashed line are the colour relations of Ref. 50. for $\log g = 4.5$ and 4.0, 
respectively. In the bottom panel, $\Delta$ means $U-V$ from Ref. 50 minus 
that from Ref. 51. The shaded region corresponds to the colour range of our sample 
stars.  }
\vspace{0mm} %% add extra space ONLY when figures/tables are "colliding"!
\end{figure}
\clearpage

\begin{table}[t]
\renewcommand{\tablename}{Supplementary Table}
\caption{Results of Monte-Carlo simulations}
\centering
\begin{tabular}{ccccc}\hline
\hline  
Simulation &  $i_{\mathrm{peak}}$ ($^{\circ}$) & $\sigma_{i}$ ($^{\circ}$) & $\alpha$ (km$^{-1}$ s) &  Confidence level (\%)\\
\hline
Case 1 & - & - & - & $6.2\pm 4.0$ \\
Case 2 &54& 1& - &  $42.1 \pm 13.2$ \\
Case 3 & - & - & 0.0006 & $10.7 \pm 4.6$\\
Case 4 & 50 & 2 & 0.0013 &  $74.1 \pm 12.8$ \\      
\hline
 \hline                  
\end{tabular}
\end{table}
\clearpage

\begin{table}[t]
\renewcommand{\tablename}{Supplementary Table}
\caption{Line list}
\centering
\begin{tabular}{cc}\hline
\hline  
Spectral lines & Spectrographs\\      
\hline
Sr {\scriptsize \textsc{II}} $\lambda 4215$ & UVES \\
Cr {\scriptsize \textsc{II}} $\lambda 4242$ & UVES\\
Cr {\scriptsize \textsc{II}} $\lambda 4261$ & UVES\\
Fe {\scriptsize \textsc{I}} $\lambda 4235$ & UVES\\
Fe {\scriptsize \textsc{I}} $\lambda 4404$ & UVES and GIRAFFE\\
Fe {\scriptsize \textsc{I}} $\lambda 4415$ & UVES and GIRAFFE\\
Ti {\scriptsize \textsc{II}} $\lambda4468$ & UVES and GIRAFFE\\
Ti {\scriptsize \textsc{II}} $\lambda 4488$ & UVES and GIRAFFE\\
Fe {\scriptsize \textsc{II}} $\lambda 4489$ & UVES and GIRAFFE\\
Fe {\scriptsize \textsc{II}} $\lambda 4491$ & UVES and GIRAFFE\\
 \hline                  
\end{tabular}
\end{table}

\end{document}